\documentclass[11pt]{article}
\usepackage[utf8]{inputenc}
\usepackage[T1]{fontenc}
\usepackage{amsmath,amssymb,amsthm,mathtools}
\usepackage{booktabs,array}
\usepackage{graphicx}
\usepackage[numbers,sort&compress]{natbib}
\usepackage[hidelinks]{hyperref}
\usepackage{geometry}
\theoremstyle{plain}
\newtheorem{theorem}{Theorem}
\newtheorem{proposition}{Proposition}
\newtheorem{corollary}{Corollary}
\newtheorem{lemma}{Lemma}
\theoremstyle{remark}
\newtheorem{remark}{Remark}
\theoremstyle{definition}
\newtheorem{definition}{Definition}
\newtheorem{assumption}{Assumption}
\newtheorem{example}{Example}

\newcommand{\E}{\mathbb{E}}
\newcommand{\Var}{\mathrm{Var}}
\newcommand{\Cov}{\mathrm{Cov}}
\newcommand{\R}{\mathbb{R}}

\newcommand{\bef}{\mathbf{f}}
\newcommand{\bh}{\mathbf{h}}
\newcommand{\bM}{\mathbf{M}}

\newcommand{\btheta}{\boldsymbol{\theta}}
\newcommand{\PMM}{\mathrm{PMM}}
\newcommand{\SLS}{\mathrm{SLSE}}
\newcommand{\OLS}{\mathrm{OLS}}
\newcommand{\gS}{g_{S}}
\newcommand{\Mg}{\bM_{g_{S}}}
\newcommand{\appref}[1]{Appendix~\ref{#1}}
\DeclareMathOperator{\tr}{tr}

\title{\bfseries Optimal Experimental Design for the Polynomial Maximization Method:\\
a Higher-Order Reserve Beyond Second-Order Design Theory}
\author{Serhii Zabolotnii\\[2pt]
\small Cherkasy State Business College, Cherkasy 18028, Ukraine\\
\small State Scientific Research Institute of Armament and Military Equipment\\
\small Testing and Certification, Cherkasy, Ukraine\\
\small Department of Cybernetics and Applied Mathematics, Uzhhorod University,\\
\small Uzhhorod 88000, Ukraine\\[2pt]
\small \texttt{zabolotnii.serhii@csbc.edu.ua}\quad ORCID: 0000-0003-0242-2234}
\date{October 2026}

\begin{document}
\maketitle
\begin{abstract}
Optimal designs for regression with asymmetric errors build on the second-order least squares estimator ($\SLS$) of Wang and Leblanc (2008), whose gain over ordinary least squares ($\OLS$) comes from the error skewness. With i.i.d.\ errors the optimally weighted $\SLS$ solves the same estimating equation as the degree-two Polynomial Maximization Method ($\PMM$), a family whose degree-$S$ location estimator has asymptotic variance $\gS\le1$ times that of the sample mean. We derive optimal designs for this family when the errors are independent with constant variance but have standardised cumulants of order three and higher that vary with the regressor (\emph{cumulant heteroscedasticity}). For a known reduction field $\gS(x)$, the $\gS$-weighted information matrix $\Mg$ obeys the Kiefer--Wolfowitz equivalence theorem. A constant field leaves the $\OLS$-optimal design unchanged; a non-constant field can change it but need not. For symmetric errors with varying kurtosis, the $\PMM_2$ and conditional-$\SLS$ designs cannot depart from the $\OLS$-optimal one, whereas a degree-three design can. A stratify-and-pool $\PMM_S$ estimator attains $N\Cov(\hat\btheta)\to\sigma^2\Mg^{-1}$ at every degree as all strata grow; its per-stratum reduction is below one only for symmetric errors with the symmetry imposed. For a quadratic model on $[0,1]$ with symmetric Beta errors of unit variance and dose-varying shape, the third-order design for the known field has $A$-efficiency $1.047$ relative to the $\OLS$-optimal design, and $1.001$ under the $D$-criterion. A two-stage procedure that estimates the field from a one-third pilot builds designs of $A$-efficiency $1.014$--$1.020$ at $N=600$--$2400$; its realised mean-squared-error ratio against the $\OLS$-optimal design ranges from $0.89$ (a loss) to $1.08$.
\end{abstract}



\section{Introduction}
\label{sec:intro}

Polynomial and fractional-polynomial regression are standard tools for \emph{dose--response} modelling and for nonlinear covariate effects in the biomedical, pharmacological and reliability sciences. In these settings the errors are often skewed, heavy-tailed or otherwise non-Gaussian; the theory below needs finite error moments of order $2S$ at degree $S$, which excludes the heaviest tails. The optimal-design problem asks where to place the design points (the doses) and how to allocate the sampling effort (the weights) so as to maximise the precision of the estimated regression coefficients. Classical optimal-design theory answers this question for $\OLS$ estimation under homoscedastic errors; $\OLS$ is then best among linear unbiased estimators whatever the error law, and Gaussianity is needed only for efficiency beyond that class. There the answer depends only on the mean structure, through the information matrix $\bM(\xi)$ and the equivalence apparatus \citep{kiefer1960equivalence,fedorov1972theory,pukelsheim2006optimal,atkinson2007optimum}. The error distribution enters only as a scalar nuisance and never moves the optimal design.

The line of work on optimal design for \emph{asymmetric} errors that we build on is based on the second-order least squares estimator ($\SLS$) of Wang and Leblanc \citep{wang2003estimation,wang2008second}, whose efficiency \citet{kim2012efficiency} studied further. This estimator augments the ordinary residual equations with their squares and so extracts information from the third moment of the error. Building on $\SLS$, Gao and Zhou \citep{gao2014optimal}, He and Yue \citep{he2018roptimality} and He, Yue and Du \citep{he2024comparing} derived optimal and robust designs whose advantage over the $\OLS$-optimal design is governed by the error skewness. He, Yue and Zheng \citep{he2026dose} recently extended this line to active-controlled dose--response studies. This is a \emph{second-order} theory. With a constant error variance its reserve is non-zero only through the third moment (the kurtosis changes its size but cannot create it), so it says nothing about symmetric departures from normality: for symmetric heavy- or light-tailed errors the skewness vanishes and $\SLS$ gains nothing over $\OLS$. When the variance also depends on the mean, the squared residuals carry information of their own, and \citet{zhang2024efficiency} let the skewness and kurtosis vary across design points in that setting.

A separate tradition, the Polynomial Maximization Method ($\PMM$) of the Cherkasy school of Kunchenko \citep{kunchenko2002polynomial}, solves the same estimation problem with a \emph{degree-graded} efficiency reserve. A degree-$S$ stochastic polynomial multiplies the asymptotic variance of the location estimator by Kunchenko's coefficient $\gS\in(0,1]$, which uses the error cumulants up to order $2S$. For $S=2$ and known scale, $g_2 = 1-\gamma_3^2/(2+\gamma_4)$; when each dose has its own estimated scale, the degree-2 location reduction is one instead (Lemma~\ref{lem:loc2}). In the design problem below $\gS$ becomes a function of the dose. A recent unification \citep{zabolotnii2026slspmm} establishes that, for linear regression with i.i.d.\ errors, optimally weighted $\SLS$ and the degree-two $\PMM$ solve the same estimating equation, and that the reserve is monotone, $1\ge g_2\ge g_3\ge\cdots$, so degrees $S\ge3$ can extract information that the second-order moment span does not contain. Here ``order'' is the polynomial degree $S$ of the estimating equation, not the order of a moment: the degree-2 ($\SLS$) equation already uses moments up to order four, but its reserve needs a non-zero third cumulant. All these statements presuppose finite error moments of order $2S$, which excludes the heaviest tails at a given degree. The method is implemented in the R package EstemPMM, and a fractional-polynomial $\PMM$ for skewed dose--response and survival data has been developed \citep{zabolotnii2026pmmfp}. The same order-two pattern, an effect carried by the third moment alone that vanishes for every symmetric law, appears in nonlinear system identification. There, identifying Volterra kernels in a Kunchenko basis orthonormal for the input law removes the excess risk that the classical Gaussian-basis cross-correlation estimator incurs under skewed input \citep{zabolotnii2026volterra}.

Both developments concern \emph{estimation} on a \emph{fixed} design. The optimal designs for asymmetric errors cited above all target the second-order $\SLS$, and we found no optimal-design analysis of the degree-$S$ $\PMM$ estimators ($S\ge3$) under cumulant heteroscedasticity. This paper asks whether the higher-order reserve changes \emph{where one should collect the data}. It builds the optimal-design theory for these estimators in a regression with independent errors of constant variance whose shape varies with the dose over a compact design region, with the resulting reduction field treated as known.

Because the field is treated as known, the optimality theory does not solve the design problem when the field must be estimated; the two-stage simulation of Section~\ref{subsec:kurt} illustrates that limitation for one pilot rule. The theory is written for the quadratic regression of the worked example. Definition~\ref{def:Mg}, Proposition~\ref{prop:adaptive} and Theorem~\ref{thm:equiv} use the regressor only through $\bef(x)\bef(x)^\top$ and hold for any regressor vector, polynomial or fractional-polynomial, with a non-singular $\bM(\xi)$; part (b) of Theorem~\ref{thm:dichotomy} is specific to the quadratic. The relation of this theory to the existing one is one of degree, not an identity of estimators. Under i.i.d.\ errors the optimally weighted $\SLS$ on which that theory rests solves the same estimating equation as the degree-2 $\PMM$ \citep[Theorem~1]{zabolotnii2026slspmm}, so its designs are designs for a degree-2 member of the family. Once the error shape varies with the dose, the two second-order estimators are no longer equivalent: the conditional $\SLS$ has its own information matrix $\mathbf{I}_2(\xi)$ (Proposition~\ref{prop:csls}), while the stratify-and-pool $\PMM_2$ has the $\OLS$ information (Lemma~\ref{lem:loc2}). The conditional-$\SLS$ design can then follow a varying skewness, but neither second-order design responds to a symmetric error field.

\paragraph{Contributions}
\begin{itemize}
  \item We define the $\gS$-weighted design-information matrix $\Mg(\xi)=\int \gS(x)^{-1}\bef(x)\bef(x)^\top d\xi(x)$ and the transformed regressor $\gS(x)^{-1/2}\bef(x)$, which recast $\PMM$ optimal design as ordinary optimal design in transformed coordinates. This item and the next two are statements about $\Mg$ for a given field and hold for every degree $S$; which estimator has $\Mg$ as its information is the subject of the fifth item.
  \item We transport the Kiefer--Wolfowitz equivalence theorem to $\Mg$, so that $D$-, $A$- and $c$-optimal $\PMM$ designs are characterised and computed by the classical multiplicative and exchange algorithms \citep{titterington1976algorithms,fedorov1972theory}.
  \item We prove a \emph{degeneracy dichotomy}. When $\gS$ is constant on the design region, the $\PMM_S$-optimal design coincides with the $\OLS$-optimal design and does not depend on the error cumulants. When $\gS$ is non-constant, which can happen only under cumulant heteroscedasticity, $\Mg$ is a pointwise reweighting of $\bM$ that need not be proportional to it, and the $\PMM_S$-optimal design can then differ from the $\OLS$-optimal one (Theorem~\ref{thm:dichotomy}(b) gives the exact condition for the quadratic model). The homoscedastic-cumulant case (constant $\gS$) is thus a degenerate boundary, on which the design is invariant to a positive scalar factor of $\bM(\xi)$.
  \item We locate the boundary between second- and third-order design (Theorem~\ref{thm:reserve}). In the symmetric class both second-order estimators keep the $\OLS$ information, however the kurtosis varies. They are the locally-adaptive $\PMM_2$ (degree-2 location fits per dose, each with its own scale, pooled by least squares), whose per-stratum location reduction is identically one (Lemma~\ref{lem:loc2}), and the global conditional $\SLS$ (one estimating equation for all doses with a common variance, weighted by the conditional second-moment matrix). A design driven by the kurtosis therefore cannot depart from the $\OLS$-optimal one at $S=2$ and can first depart at $S=3$, when $g_3$ is non-constant on the support. A varying skewness is the exception: the global conditional $\SLS$, which we analyse as the second-order comparator, uses the common error variance and responds to it (Proposition~\ref{prop:csls}).
  \item We give a locally-adaptive $\PMM_S$ estimator whose asymptotic covariance is $\sigma^2\Mg(\xi)^{-1}$ under a per-stratum limit (Assumption~\ref{ass:stratum}; Proposition~\ref{prop:adaptive}, proved in full in \appref{app:prop1}), so that, for a known field, it attains the covariance for which the design is optimal. This is an oracle-field result; plug-in estimation of the field and the two-stage simulation are treated separately.
  \item We prove the per-stratum limit at every degree (Lemma~\ref{lem:gS}). With no restriction on the error shape (neither symmetry nor a cumulant constraint) the per-stratum estimator is the stratum mean at every degree ($\gS\equiv1$; Lemma~\ref{lem:loc2} is the case $S=2$). With symmetric errors and the symmetry imposed it attains an odd-power reduction $\gS\le1$, which for $S=3$ we derive in closed form (\appref{app:g3}). The stratify-and-pool estimator therefore realises a non-trivial field only in the symmetric class; for asymmetric errors its information is the $\OLS$ one, although the design results for a given field still apply. We also discuss the sample size the reserve needs to survive plug-in cumulant estimation.
  \item We bound the two departures from the oracle setting. A plug-in design is asymptotically efficient when the field estimate is consistent on the whole grid and the pilot fraction vanishes, and otherwise converges to an explicit limit (Proposition~\ref{prop:plugin}). Under a local asymmetry the degree-3 stratum root acquires a bias, and the stratum keeps a gain over its mean only while its skewness is below one standard error of its sample skewness in absolute value (Proposition~\ref{prop:asym}, Corollary~\ref{cor:asym}).
  \item We illustrate the theory on a quadratic dose--response model in EstemPMM. In a symmetric, kurtosis-heteroscedastic \textsc{kurt} regime the third-order design separates. For a known field its analytic $A$-efficiency relative to the $\OLS$-optimal design is $1.047$ ($+4.7\%$), and a Monte Carlo study of the locally-adaptive estimator with the field known gives realised gains of $1.035$ to $1.061$ at $N=600$ to $2400$. A two-stage design that estimates the field from a one-third pilot captures only part of the gain at $N\le2400$; this is one pilot rule in one case study, not a general finding about plug-in designs. In a \textsc{skew} contrast the locally-adaptive $\PMM_2$ does not separate while the global conditional $\SLS$ does ($+2.5\%$ analytic $A$-gain). The second-order channel therefore reads a skewness field and is blind only to a symmetric one.
\end{itemize}

\paragraph{Positioning}
This is an \emph{experimental-design} paper, and it builds on a companion estimation preprint without repeating its results. The unification paper \citep{zabolotnii2026slspmm} is an estimation result on a fixed design under i.i.d.\ errors, and it proves that the $\SLS$ slope efficiency is then design-invariant. We take from it the $\SLS=\PMM_2$ identity of estimating equations and the degree ordering $1\ge g_2\ge g_3\ge\cdots$, both as background. That preprint does not treat experimental design. The results that our theorems rest on are proved here: the unit second-order location reduction (Lemma~\ref{lem:loc2}), the conditional-$\SLS$ information (Proposition~\ref{prop:csls}), whose part (b) is the pointwise form of the companion's symmetric-blindness result, and the third-order location reduction with co-estimated scale and shape (Lemma~\ref{lem:g3}).

The new object here is design optimisation under cumulant heteroscedasticity: error cumulants of order three and higher vary with the dose while the variance stays constant, and the resulting reduction field is treated as known. The integer degree-3 basis of our study is the estimator inside that design problem.

Relative to the $\SLS$ optimal-design line \citep{gao2014optimal,bose2015optimal,he2018roptimality,he2024comparing,he2026dose,zhang2024efficiency}, reviewed in \S\ref{sec:bg-asymm}, we extend the foundation rather than compete with it. Under i.i.d.\ errors those designs are designs for the estimator that solves the degree-2 $\PMM$ equation. At constant variance those designs keep one lever, the skewness, which a varying field can make design-relevant (Proposition~\ref{prop:csls}), and they cannot reach the symmetric, kurtosis-heteroscedastic regime. The analytic efficiency gain is small. The contribution is structural: a new design problem, its relation to the second-order theory, and, for symmetric errors with varying kurtosis, a separation regime that opens only at degree $S\ge3$.

\section{Background}
\label{sec:background}

\subsection{The Polynomial Maximization Method and the reduction coefficient \texorpdfstring{$\gS$}{gS}}
\label{sec:bg-pmm}

Consider the estimation of a location parameter $\mu$ from i.i.d.\ observations $z_i=\mu+\varepsilon_i$ with zero-mean errors of variance $\sigma^2$ and standardised cumulants $\gamma_3,\gamma_4,\dots$ (skewness, excess kurtosis, and higher orders). The \emph{Polynomial Maximization Method} ($\PMM$) of Kunchenko \citep{kunchenko2002polynomial} estimates $\mu$ by maximising a degree-$S$ stochastic polynomial in the data, i.e.\ by solving an estimating equation built from the empirical power moments up to order $2S$. The method has a \emph{degree-graded efficiency reserve}. For a degree-$S$ polynomial, and with the scale parameter $\sigma^2$ known, the asymptotic variance of the $\PMM$ location estimator is
\begin{equation}
\Var(\hat\mu_{\PMM_S}) \;=\; \frac{\sigma^2\, \gS}{N}\bigl(1+o(1)\bigr),
\qquad \gS=\gS(\gamma_3,\gamma_4,\dots)\in(0,1],
\label{eq:gS-var}
\end{equation}
where $\gS$ is Kunchenko's \emph{variance-reduction coefficient}, a closed-form rational function of the standardised cumulants up to order $2S$ \citep{kunchenko2002polynomial,zabolotnii2026slspmm}. The factor $\gS\le 1$ measures how much the higher cumulants are used relative to the sample mean: $\gS=1$ recovers ordinary least squares (no reserve), while $\gS<1$ is a strict variance reduction. For degree $S=2$ the coefficient has the explicit form
\begin{equation}
g_2 \;=\; 1-\frac{\gamma_3^{2}}{2+\gamma_4},
\label{eq:g2}
\end{equation}
which is driven by the error \emph{skewness} $\gamma_3$ and equal to one whenever the error law is symmetric. Whether this degree-2 reserve is \emph{realised in the location} parameter depends on the estimation setup. It is realised when the scale is known. It is not realised when the scale is co-estimated and the mean is constant, which is the case of our design strata (Lemma~\ref{lem:loc2}). The method is implemented in the R package EstemPMM (see Code availability), which we use as the estimation engine throughout.

\subsection{SLS as degree-2 PMM and the degree ordering}
\label{sec:bg-slse}

The optimal-design literature has approached estimation under non-Gaussian errors through the \emph{second-order least squares estimator} ($\SLS$) \citep{wang2003estimation,wang2008second,kim2012efficiency}. It augments the ordinary least-squares estimating equations with the second-moment (squared-residual) equations and gains efficiency over $\OLS$ when the errors are asymmetric. The recent unification \citep[Theorem~1]{zabolotnii2026slspmm} shows that, for linear regression with i.i.d.\ errors, the optimally weighted $\SLS$ and the degree-2 $\PMM$ estimator solve the same population normal system and share one influence function, so they are asymptotically equivalent, though not identical in finite samples. Their asymptotic slope variance is $g_2$ times that of $\OLS$, an asymptotic relative efficiency of $1/g_2$, with $g_2$ given by \eqref{eq:g2}. The identity is one of estimating equations, and it needs a design-free error variance and the optimal combination. Write $m(e)=(e,\,e^2-\sigma^2)^\top$, $F_2(x)$ for its conditional covariance matrix at $x$ and $b=(1,0)^\top$; the optimal combination of the two residual equations is then $F_2(x)^{-1}b$ (\eqref{eq:csls-score} below). When the conditional variance or this combination varies with the regressor, the two become different estimators: the conditional $\SLS$ weights each observation by its conditional second-moment matrix, and the unconditional $\PMM_2$ does not \citep[Proposition~1]{zabolotnii2026slspmm}. Cumulant heteroscedasticity, the regime of this paper, is such a case once the skewness varies, and Section~\ref{sec:dichotomy} keeps the two apart. \citet{zabolotnii2026slspmm} also proves that the reduction coefficients are monotone in the polynomial degree,
\begin{equation}
1 \;\ge\; g_2 \;\ge\; g_3 \;\ge\; \cdots \;\ge\; g_S \;\ge\; \cdots \;>\;0,
\label{eq:degree-ordering}
\end{equation}
so that an increment in degree cannot reduce efficiency (equality is possible, for example $g_2=1$ for a symmetric law), using cumulants of successively higher order. The ordering concerns the known-scale coefficients; for the realised per-stratum reductions that enter the design below we rely only on Lemmas~\ref{lem:loc2} and~\ref{lem:g3}, which are proved here. The reserve is a property of the moment span rather than of the algorithm. Any regular estimator that is asymptotically efficient on the degree-$S$ moment span with optimal weighting attains the same asymptotic factor $g_S$ \citep[Section~4.4]{zabolotnii2026slspmm}, though not necessarily the same finite-sample behaviour. A two-step generalised method of moments on the same conditions \citep{hansen1982large} is one example; $\PMM$ supplies the closed-form normal system and the graded family. The consequence we use most is that the degree-2 ($\SLS$) reserve acts only through the \emph{odd} cumulant $\gamma_3$. By \eqref{eq:g2}, $g_2\equiv1$ as soon as $\gamma_3=0$, whatever the even cumulants, and for a symmetric law the optimal second-order combination places zero weight on the squared residual \citep[Proposition~2]{zabolotnii2026slspmm}. For degrees $S\ge3$ the reserve also responds to the \emph{even} cumulants $\gamma_4,\gamma_6,\dots$. We take the ordering \eqref{eq:degree-ordering} as an estimation-side input and do not re-derive it. The present paper asks a different question: how this graded reserve interacts with the choice of experimental design.

\subsection{Optimal design and the equivalence theorem}
\label{sec:bg-design}

For the regression $y_i=\bef(x_i)^\top\btheta+\varepsilon_i$ on a design region $\mathcal{X}$, an (approximate) design $\xi$ is a probability measure with support points $x_k$ and weights $w_k\ge0$, $\sum_k w_k=1$. Under Gaussian errors and $\OLS$ the asymptotic covariance of $\hat\btheta$ is proportional to $\bM(\xi)^{-1}$, where
\begin{equation}
\bM(\xi)=\int_{\mathcal{X}}\bef(x)\bef(x)^\top\,d\xi(x)=\sum_k w_k\,\bef(x_k)\bef(x_k)^\top,
\label{eq:fisher-ols}
\end{equation}
is the classical information matrix \citep{atkinson2007optimum,pukelsheim2006optimal}. Optimal design chooses $\xi$ to minimise a scalar criterion $\Psi$ of $\bM(\xi)^{-1}$: $D$-optimality maximises $\det\bM(\xi)$, $A$-optimality minimises $\tr\bM(\xi)^{-1}$ (the total coefficient variance), and $c$-optimality targets a linear functional. The cornerstone of the theory is the Kiefer--Wolfowitz \emph{general equivalence theorem} \citep{kiefer1960equivalence}: a design $\xi^*$ is $D$-optimal if and only if
\begin{equation}
\max_{x\in\mathcal{X}}\ \bef(x)^\top\bM(\xi^*)^{-1}\bef(x)=p,
\label{eq:kw-equivalence}
\end{equation}
with $p=\dim\btheta$, the maximum being attained at the support points of $\xi^*$; analogous variational characterisations hold for $A$- and $c$-optimality \citep{pukelsheim2006optimal,atkinson2007optimum}. The theorem makes optimality checkable at a candidate design and underlies the multiplicative and exchange algorithms used to compute optimal designs \citep{fedorov1972theory,titterington1976algorithms,silvey1978algorithm,yu2010monotonic}. A property we use repeatedly is the \emph{scale invariance} of every standard criterion: replacing $\bM(\xi)$ by $c\,\bM(\xi)$ for any constant $c>0$ leaves the $\Psi$-optimal design unchanged.

\subsection{The asymmetric-error optimal-design line}
\label{sec:bg-asymm}

The classical theory for the homoscedastic linear model uses the $\OLS$ estimator \citep{pukelsheim2006optimal,atkinson2007optimum}. The error distribution then enters the covariance $\sigma^2\bM(\xi)^{-1}$ only through the scalar $\sigma^2$, which does not affect the optimal design. The line of work that relaxed this for \emph{asymmetric} errors, meaning errors with non-zero skewness, is built on the second-order $\SLS$ \citep{wang2003estimation,wang2008second}. \citet{gao2014optimal} introduced design criteria for the $\SLS$ and showed that its advantage over $\OLS$ is governed by the error skewness, and \citet{gao2017doptimal} studied its $D$-optimal designs. \citet{bose2015optimal} developed the approximate theory, in which the $\SLS$ information matrix is nonlinear in the design measure; \citet{huda2018string} and \citet{yeh2021properties} derived structural properties of the resulting designs, and \citet{he2018multifactor} extended them to multi-factor nonlinear models. \citet{he2018roptimality} developed $R$-optimality for the same estimator, \citet{he2024comparing} constructed $\SLS$-based designs for comparing regression curves, and \citet{he2026dose} treated active-controlled dose--response studies. These works vary the criterion, the model and the treatment of the nonlinear information, but in each of them the optimised estimator is the second-order one and the error law, apart from its mean, does not change with the regressor. By the identification of \S\ref{sec:bg-slse}, they therefore optimise designs for an estimator that solves the degree-2 $\PMM$ equation. Likelihood-based designs for a skewed error family \citep{zhou2024skewnormal} likewise keep the error shape fixed across the design region.

Designs under an error law that changes with the regressor have mostly been studied through the variance. Heteroscedastic optimal designs weight the information matrix by an inverse variance function \citep{wong1993heteroscedastic,atkinson1995heteroscedastic}, and designs for the maximum quasi-likelihood estimator do so through a mean--variance relation \citep{shen2016mqle}. A weighting of $\bef(x)\bef(x)^\top$ by a pointwise functional of the error law also appears in optimal designs for quantile regression \citep{dette2012quantile}. Closest to the present work, \citet{zhang2024efficiency} construct locally $D$-optimal designs for an oracle $\SLS$ whose weights use the skewness and kurtosis at each design point, under a mean--variance relation, and compare them with designs for the quasi-likelihood and Gaussian-likelihood estimators. Their designs respond to a varying error shape, so second-order design theory is not confined to a constant law. Two features separate their problem from ours. In their model the variance depends on the mean, so the squared-residual equation carries information on the regression parameters whatever the skewness; here the variance is constant, and for symmetric errors that channel is closed (Theorem~\ref{thm:reserve}(i)). And their estimator has degree two, whereas the separation studied here opens only at degree $S\ge3$. The conditional $\SLS$ of Proposition~\ref{prop:csls} is, in this sense, the constant-variance analogue of their oracle estimator.

To our knowledge the design problem has not been posed for the higher-order ($S\ge3$) estimators, whose variance-reduction factor $\gS$ in \eqref{eq:degree-ordering} also depends on the even cumulants $\gamma_4,\gamma_6,\dots$. For $S=2$, by contrast, the coefficient \eqref{eq:g2} is driven by the third cumulant $\gamma_3$, and $\gamma_4$ affects it only when $\gamma_3\ne0$. The present paper poses this problem for $S\ge3$ under dose-varying higher cumulants, with the field known; for plug-in designs we give the limit of their efficiency (Proposition~\ref{prop:plugin}) and a simulation. Suppose the errors are symmetric at every dose, with a common variance and a kurtosis that may vary. Then the two second-order designs studied here cannot respond to the error field. Symmetry removes the third cumulant, through which alone the second-order reserve acts, and both second-order information matrices stay proportional to $\bM$ however the even cumulants vary (Theorem~\ref{thm:reserve}(i)), so neither the optimal design nor its covariance differs from that of $\OLS$. A third-order design, by contrast, can respond, because $g_3$ depends on $\gamma_4$ and $\gamma_6$ and so varies with the dose. Within each dose stratum, estimating a location under a symmetric law of unknown shape is the classical problem of adaptive estimation, in which the Fisher bound can be attained without knowing the law \citep{stone1975adaptive,bickel1982adaptive}. The degree-$S$ $\PMM$ estimator is a fixed-degree polynomial solution of that problem: it attains $\gS$, not the Fisher bound.

\section{The PMM design-information matrix}
\label{sec:designinfo}

We retain the regression $y_i=\bef(x_i)^\top\btheta+\varepsilon_i$ with quadratic mean
$\bef(x)=(1,x,x^2)^\top$, $\btheta\in\R^{3}$, $p=3$, and errors that are independent with
$\E(\varepsilon_i\mid x_i)=0$ and $\Var(\varepsilon_i\mid x_i)=\sigma^2$. What is allowed to move is the
\emph{shape} of the conditional error law, so the errors are independent but in general not
identically distributed. We say the errors exhibit \emph{cumulant heteroscedasticity}
when the variance $\sigma^2$ is constant across the design region $\mathcal{X}=[0,1]$ while
one or more of the standardised cumulants---the skewness $\gamma_3(x)$, the excess kurtosis
$\gamma_4(x)$, and those of higher order---vary with the regressor $x$. The classical
situation in which every $\gamma_k$ is constant we call \emph{cumulant-homoscedastic}; it is
the i.i.d.\ setting of the second-order optimal-design line. Cumulant heteroscedasticity is the
central notion of this paper. The variance carries no spatial information, whereas the
higher cumulants do, and which of them a design can use depends on the degree of the
estimator. In the symmetric class a second-order estimator has nothing to read, because $g_2=1$
whenever $\gamma_3=0$, while a third-order one reads a varying kurtosis; outside it, a second-order
estimator reads the kurtosis only together with a non-zero skewness (Section~\ref{sec:dichotomy}).

Kunchenko's variance-reduction coefficient turns this spatial variation of shape into a
spatial variation of attainable precision. With known scale, a degree-$S$ stochastic
polynomial at dose $x$ attains Kunchenko's coefficient evaluated at the local cumulants,
$\gS\big(\gamma_3(x),\gamma_4(x),\dots\big)\in(0,1]$
\citep{kunchenko2002polynomial,zabolotnii2026slspmm}; for example
$g_2=1-\gamma_3^2/(2+\gamma_4)$ evaluated pointwise. Such coefficients have a projection reading: the fraction of the score captured by a finite polynomial span is the estimator's asymptotic relative efficiency against maximum likelihood \citep{zabolotnii2026lsu}; here $\gS$ is measured against the stratum mean instead. The \emph{reduction field} $\gS(x)$ that enters the
design is the realised version of this coefficient, defined at the end of this paragraph. Under cumulant homoscedasticity $\gS$ is
a constant; under cumulant heteroscedasticity it may vary on $\mathcal{X}$, and whether it does
depends on the degree $S$ and on which cumulants vary (Section~\ref{sec:dichotomy}). We assume
that at every $x$ the conditional error law has finite moments of order $2S$ and is not
concentrated on $S$ or fewer points, so that $0<\gS(x)\le1$. These moment conditions suffice for the per-stratum limits proved below (Lemmas~\ref{lem:g3} and~\ref{lem:gS}), whose plug-in step uses only the consistency of the sample moments, not their asymptotic normality. The estimated coefficients enter in directions orthogonal to the location score, so their error does not affect the first-order limit. For the equivalence theorem on the
continuous region we also assume that $\gS$ is continuous on the compact $\mathcal{X}$, hence bounded
away from zero; then $\bh$ is continuous, the maxima below are attained, and optimal approximate designs
exist \citep{pukelsheim2006optimal}. Throughout, $\gS(x)$ denotes the
reduction that the per-stratum degree-$S$ location estimator actually achieves, i.e.\ the ratio
of its asymptotic variance to that of the stratum mean, with the scale estimated within the
stratum. With known scale this is Kunchenko's coefficient evaluated at the local cumulants; with a
co-estimated scale it can be closer to one, and for $S=2$ it equals one (Lemma~\ref{lem:loc2}). Its value depends on the shape restriction the stratum imposes: with none, $\gS\equiv1$ at every degree; with symmetry imposed on a symmetric law, $\gS$ is the odd-power coefficient of Lemma~\ref{lem:gS} (Assumption~\ref{ass:stratum} below). The third-order field of the numerical study is of the second kind. Where the
known-scale coefficient \eqref{eq:g2} is meant instead, the text says so. The
design problem is to choose where to place doses and how to allocate samples so as to obtain
the most precision \emph{given} this field.

\begin{definition}[PMM design-information matrix]
\label{def:Mg}
Fix a degree $S$ and an approximate design $\xi$ with support points $x_k$ and weights
$w_k\ge0$, $\sum_k w_k=1$. The \emph{$\gS$-weighted ($\PMM_S$ design-)information matrix} of
$\xi$ is
\begin{equation}
\label{eq:Mg}
\begin{aligned}
\Mg(\xi)
&:=\int_{\mathcal{X}}\gS(x)^{-1}\,\bef(x)\bef(x)^\top\,d\xi(x)
=\sum_k \frac{w_k}{\gS(x_k)}\,\bef(x_k)\bef(x_k)^\top\\
&\phantom{:}=\int_{\mathcal{X}}\bh(x)\bh(x)^\top\,d\xi(x),
\end{aligned}
\end{equation}
where
\[
\bh(x):=\gS(x)^{-1/2}\,\bef(x)
\]
is the \emph{transformed regressor}. Equivalently, $\Mg$ is the ordinary information matrix of
the design $\xi$ read in the regressors $\bh$ rather than $\bef$.
\end{definition}

The matrix in \eqref{eq:Mg} is the main tool of the paper. It reweights each support
point by the inverse of the local reserve $\gS(x_k)$. Strata where the degree-$S$ polynomial
buys a large variance reduction (small $\gS$) receive a large information weight. Strata
where it buys nothing ($\gS\equiv1$) are weighted as in the classical least-squares
information matrix $\bM(\xi)=\int\bef\bef^\top\,d\xi$. In particular, when $\gS(x)\equiv1$ on
the support, the transformed regressor reduces to $\bh\equiv\bef$ and
$\Mg(\xi)=\bM(\xi)$, the ordinary $\OLS$ information. Everything that is specific to the
$\PMM$ design problem comes from the variation of $\gS$. The matrix $\Mg$ is the information of the
stratify-then-pool procedure of Proposition~\ref{prop:adaptive} below (per-stratum $\PMM_S$ location
fits combined by generalised least squares), not of a $\PMM$ regression fitted to all observations at
once; the global conditional $\SLS$ of Section~\ref{sec:dichotomy} has a different information matrix.

We now show that $\Mg$ is the asymptotic precision of an implementable estimator, not
only a formal reweighting.

\begin{assumption}[Per-stratum limit]
\label{ass:stratum}
At every support point $x_k$, the per-stratum degree-$S$ $\PMM$ location estimator $\hat\mu_k$ of
$\mu_k=\bef(x_k)^\top\btheta$, computed from the $n_k$ observations at $x_k$ with the scale and shape
estimated within the stratum (the estimating equations, and the shape restriction, none or symmetry, are those of
Lemma~\ref{lem:gS}), satisfies
\[
\sqrt{n_k}\,(\hat\mu_k-\mu_k)\xrightarrow{\,d\,}N\!\big(0,\ \sigma^2\,\gS(x_k)\big),\qquad n_k\to\infty,
\]
with $\gS(x_k)\in(0,1]$ the reduction field of Definition~\ref{def:Mg}.
\end{assumption}

Assumption~\ref{ass:stratum} holds at every degree $S$, under the standing moment and support conditions
(Lemma~\ref{lem:gS} in \appref{app:gS}). Which reduction it delivers depends on what the stratum is allowed to
know about the shape. With the shape unrestricted, the per-stratum degree-$S$ system is exactly identified and
returns the stratum mean, so $\gS(x_k)=1$ for every law and every $S$; for $S=2$ this is Lemma~\ref{lem:loc2}
below. When the conditional law is symmetric and the estimator imposes that symmetry, the reduction is the
odd-power coefficient $\gS=1/(m_2\,\mathbf{b}^\top\mathbf{F}^{-1}\mathbf{b})\le1$ of \eqref{eq:gS-odd}, which for
$S=3$ is the closed form of Lemma~\ref{lem:g3} in \appref{app:g3} ($g_3=1$ when $\gamma_4=0$). A
stratify-and-pool estimator therefore realises a reduction only through the symmetry restriction: an
asymmetric stratum with a free shape realises none at any degree, which is the efficiency bound for a mean
in a nonparametric model \citep[Chapter~25]{vandervaart1998asymptotic}.

\begin{proposition}[Asymptotics of the locally-adaptive $\PMM_S$ estimator]
\label{prop:adaptive}
Stratify the $N$ observations by support point and write $n_k=Nw_k$. For each stratum estimate
the local mean $\mu_k=\bef(x_k)^\top\btheta$ by the per-stratum degree-$S$ $\PMM$ location
estimator $\hat\mu_k$ of Assumption~\ref{ass:stratum}. Combine the $\{\hat\mu_k\}$ by inverse-variance-weighted (generalised) least squares with
weights proportional to $n_k/\gS(x_k)$, producing $\hat\btheta$. Suppose that Assumption~\ref{ass:stratum} holds, that $\xi$ has finitely
many support points with fixed weights $w_k>0$ and a non-singular $\bM(\xi)$, that
$0<\gS(x_k)\le1$ at every support point (the standing assumption above), that $n_k/N\to w_k$,
that the strata are independent, and that $N\to\infty$. Then
\[
\sqrt{N}\,(\hat\btheta-\btheta)\xrightarrow{\,d\,}
N\!\big(0,\ \sigma^2\,\Mg(\xi)^{-1}\big).
\]
The same limit holds when $\gS(x_k)$ in the weights is replaced by any strictly positive estimate that
is consistent for it as every $n_k\to\infty$ with the support fixed; the estimate may use the same
stratum data (\appref{app:prop1}). No other linear combination of the $\hat\mu_k$ that returns $\btheta$ whenever
$\hat\mu_k=\mu_k$ for all $k$ has a smaller limiting
covariance in the Loewner order; this is an asymptotic Gauss--Markov statement within the class of linear
pooled estimators whose coefficient matrix converges in probability to a deterministic matrix.
\end{proposition}

\begin{proof}
Since the stratum means satisfy $\mu_k=\bef(x_k)^\top\btheta$ exactly, $\hat\btheta-\btheta$ is a linear
map of $(\hat\mu_k-\mu_k)_k$ whose matrix converges to $\Mg(\xi)^{-1}\mathbf{F}^\top\mathbf{W}$, with
$\mathbf{F}$ the matrix of rows $\bef(x_k)^\top$ and $\mathbf{W}=\mathrm{diag}\{w_k/\gS(x_k)\}$. The
independent per-stratum limits give $\sqrt N(\hat\mu_k-\mu_k)_k\to N(\mathbf{0},\sigma^2\mathbf{W}^{-1})$, and
$\mathbf{F}^\top\mathbf{W}\mathbf{F}=\Mg(\xi)$ by \eqref{eq:Mg}. The full argument, with the optimality of the
weights, is in \appref{app:prop1}.
\end{proof}

Assumption~\ref{ass:stratum} is what links $\Mg$ to an implementable estimator. The equivalence theorem
and the dichotomies of Section~\ref{sec:dichotomy} are statements about $\Mg$ for a given field and do
not use it. The numerical study uses two of its cases: $S=2$ (Lemma~\ref{lem:loc2}) and $S=3$ under
symmetry, whose closed form of $g_3$ (\appref{app:g3}) the calibration of Section~\ref{sec:numerical}
approximates.

Proposition~\ref{prop:adaptive} shows that $\sigma^2\Mg(\xi)^{-1}$ is the asymptotic covariance
of an explicit, locally-adaptive $\PMM_S$ procedure: estimate each stratum's location
by the higher-order method, then pool with optimal weights. Any standard
design criterion $\Psi$ applied to $\Mg(\xi)$ therefore optimises the large-sample precision of
that estimator, and the optimal-design machinery carries over through the transformed
regressor.

\begin{theorem}[Transported Kiefer--Wolfowitz equivalence]
\label{thm:equiv}
Because $\Mg(\xi)=\int\bh\,\bh^\top\,d\xi$ with $\bh(x)=\gS(x)^{-1/2}\bef(x)$,
a $\Psi$-optimal design for $\Mg$ is exactly the ordinary $\Psi$-optimal design in the
transformed regressors $\bh$. The full equivalence apparatus applies verbatim
\citep{kiefer1960equivalence,pukelsheim2006optimal,atkinson2007optimum}. In particular, a design
$\xi^\ast$ with non-singular $\Mg(\xi^\ast)$ (equivalently, non-singular $\bM(\xi^\ast)$, since $\gS>0$) is $D$-optimal for $\Mg$ if and only if
\[
\max_{x\in\mathcal{X}}\ \bh(x)^\top\,\Mg(\xi^\ast)^{-1}\,\bh(x)=p,
\]
with equality on the support of $\xi^\ast$, where $p=3$; equivalently, in original
coordinates,
\[
\max_{x}\,\gS(x)^{-1}\,\bef(x)^\top\Mg(\xi^\ast)^{-1}\bef(x)=p .
\]
\end{theorem}

The theorem is a direct corollary of the classical one and requires no separate proof; the new content is the matrix $\Mg$ and its realisation by an estimator (Proposition~\ref{prop:adaptive}, Lemma~\ref{lem:gS}). Substituting $\bh$ for $\bef$ turns $\Mg$ into a
standard information matrix, so the Kiefer--Wolfowitz general equivalence theorem and its
$\Psi$-optimality variants hold without modification, and the standardised-variance function is
read off the transformed regressor. Optimal designs for $\Mg$ are computed by the
classical multiplicative and exchange algorithms run on $\bh(x)=\gS(x)^{-1/2}\bef(x)$
\citep{titterington1976algorithms,fedorov1972theory}. The reduction field $\gS$ enters only
through this pointwise rescaling of the regressors; everything else is inherited from
classical design theory. The next section settles \emph{when} this rescaling moves the
optimal design and when it cannot.

\begin{remark}[Number of strata]
\label{rem:strata}
Assumption~\ref{ass:stratum} and Proposition~\ref{prop:adaptive} ask every stratum to grow. For the designs
that matter this is not restrictive. Because $\Mg$ is an ordinary information matrix in $\bh$, an optimal
design can be chosen with at most $p(p+1)/2$ support points \citep{pukelsheim2006optimal}, six for the
quadratic, and the $A$- and $D$-optimal designs of Section~\ref{sec:numerical} have three, so each stratum
receives a fixed fraction of $N$. A design spread over many doses with few observations at each falls outside
Proposition~\ref{prop:adaptive}; for such a design the free-shape stratum mean ($\gS\equiv1$) is the safe fit.
\end{remark}

\begin{remark}[A varying variance]
\label{rem:variance}
The constant variance is used by the conditional $\SLS$ of Section~\ref{sec:dichotomy}, not by the
stratify-and-pool estimator. There each stratum estimates its own scale, and the reductions of
Lemmas~\ref{lem:g3} and~\ref{lem:gS} depend on the standardised moments only. If
$\Var(\varepsilon\mid x)=\sigma^2(x)$ with $\sigma^2$ continuous and positive, the proof of
Proposition~\ref{prop:adaptive} therefore goes through with the weights $n_k/\{\sigma^2(x_k)\gS(x_k)\}$, or with
consistent per-stratum estimates of $\sigma^2(x_k)$ in them, and gives the limit
$N\!\big(0,\mathbf{M}_{\sigma,g}(\xi)^{-1}\big)$ with
$\mathbf{M}_{\sigma,g}(\xi)=\int\sigma(x)^{-2}\gS(x)^{-1}\bef(x)\bef(x)^\top d\xi(x)$. The equivalence theorem then
holds in $\bh(x)=\bef(x)/\{\sigma(x)\gS(x)^{1/2}\}$, and the dichotomy of Theorem~\ref{thm:dichotomy} holds with the
classical heteroscedastic design, optimal for the efficiency function $\sigma^{-2}(x)$
\citep{wong1993heteroscedastic,atkinson1995heteroscedastic}, in place of the $\OLS$ design: a constant
$\gS$ leaves that design unchanged, and a non-constant one can move it (the proof of part (b) is unchanged with
the weights $w_k/\sigma^2(x_k)$). We did not run the numerical study under a varying variance.
\end{remark}

The optimality theory treats the field as known. When it is estimated, the design is built on the estimate
and judged under the true field. The next result states when such a plug-in design is asymptotically
efficient and what its limit is when it is not. It concerns designs on a finite grid, as computed in
Section~\ref{sec:numerical}. For a design $\xi$ and a field $g$ write $\mathrm{eff}_g(\xi)$ for its
$A$-efficiency $\min_\eta\tr\bM_g(\eta)^{-1}/\tr\bM_g(\xi)^{-1}$ or its $D$-efficiency
$\{\det\bM_g(\xi)/\max_\eta\det\bM_g(\eta)\}^{1/p}$, set to zero at a singular $\bM_g(\xi)$.

\begin{proposition}[Plug-in designs]
\label{prop:plugin}
Let the designs be probability vectors on a finite grid $G\subset\mathcal{X}$ that supports a design with
non-singular $\bM$, let $\Psi$ be the $A$- or the $D$-criterion, and let $g$ and $\tilde g$ be positive on $G$.
For a positive field $u$ on $G$ the information is $\bM_u(\xi)=\sum_{x\in G}\xi(x)\,u(x)^{-1}\bef(x)\bef(x)^\top$, as in
\eqref{eq:Mg}. Let $\hat g_N$ be positive random fields on $G$ with $\max_{G}|\hat g_N/\tilde g-1|\to0$ in probability, and let
$\hat\xi_N$ be $\Psi$-optimal for $\bM_{\hat g_N}$.
\begin{enumerate}
\item[(a)] If $\tilde g=g$, then $\mathrm{eff}_g(\hat\xi_N)\to1$ in probability.
\item[(b)] If the $\Psi$-optimal design $\tilde\xi$ for $\bM_{\tilde g}$ is unique, then $\hat\xi_N\to\tilde\xi$ and
$\mathrm{eff}_g(\hat\xi_N)\to\mathrm{eff}_g(\tilde\xi)$ in probability.
\item[(c)] For a pilot design $\xi_0$ and a pilot fraction $\lambda_N\in[0,1)$ let
$\xi_N=\lambda_N\xi_0+(1-\lambda_N)\,\mathbf{v}_N/\|\mathbf{v}_N\|_1$ with $\mathbf{v}_N=(\hat\xi_N-\lambda_N\xi_0)_+$, the
allocation of the remaining mass in proportion to the positive deficits. Then
$\|\xi_N-\hat\xi_N\|_1\le3\lambda_N$. If $\lambda_N\to0$, (a) and (b) hold with $\xi_N$ in place of $\hat\xi_N$. If
$\lambda_N\equiv\lambda$ and $\tilde\xi$ is unique, $\mathrm{eff}_g(\xi_N)$ converges in probability to the efficiency of
the same construction applied to $\tilde\xi$.
\end{enumerate}
\end{proposition}

\begin{proof}
Put $\varepsilon_N=\max_G|\hat g_N/\tilde g-1|$ and work on the event $\varepsilon_N<\tfrac12$, whose probability
tends to one. Since $\tilde g/\hat g_N\in[(1+\varepsilon_N)^{-1},(1-\varepsilon_N)^{-1}]$ on $G$,
$(1+\varepsilon_N)^{-1}\bM_{\tilde g}(\xi)\preceq\bM_{\hat g_N}(\xi)\preceq(1-\varepsilon_N)^{-1}\bM_{\tilde g}(\xi)$ in the
Loewner order for every design $\xi$. The two matrices are therefore singular together. Hence, uniformly over designs with a non-singular $\bM(\xi)$,
$(1-\varepsilon_N)\tr\bM_{\tilde g}(\xi)^{-1}\le\tr\bM_{\hat g_N}(\xi)^{-1}\le(1+\varepsilon_N)\tr\bM_{\tilde g}(\xi)^{-1}$, and
$|\log\det\bM_{\hat g_N}(\xi)-\log\det\bM_{\tilde g}(\xi)|\le p\,|\log(1-\varepsilon_N)|$; optimal designs are non-singular.
For any $\Psi$-optimal $\tilde\xi$ for $\tilde g$, the optimality of $\hat\xi_N$ for $\hat g_N$ gives, for the $A$-criterion,
\[
\tr\bM_{\tilde g}(\hat\xi_N)^{-1}\le\frac{\tr\bM_{\hat g_N}(\hat\xi_N)^{-1}}{1-\varepsilon_N}
\le\frac{\tr\bM_{\hat g_N}(\tilde\xi)^{-1}}{1-\varepsilon_N}\le\frac{1+\varepsilon_N}{1-\varepsilon_N}\,\tr\bM_{\tilde g}(\tilde\xi)^{-1},
\]
and the same chain with an additive $2p|\log(1-\varepsilon_N)|$ for the $D$-criterion. So
$\mathrm{eff}_{\tilde g}(\hat\xi_N)\to1$ in probability, which is (a) when $\tilde g=g$.
(b) The function $\mathrm{eff}_{\tilde g}$ is continuous on the compact simplex of designs on $G$ and equals one
only at $\tilde\xi$, so for every neighbourhood $U$ of $\tilde\xi$ it is bounded away from one outside $U$; hence
$\Pr(\hat\xi_N\notin U)\to0$. The function $\mathrm{eff}_g$ is continuous as well, which gives the second claim.
(c) For non-negative numbers $|(a-b)_+-a|\le b$, so $\|\mathbf{v}_N-\hat\xi_N\|_1\le\lambda_N$ and
$\|\mathbf{v}_N\|_1\in[1-\lambda_N,1]$. The triangle inequality through $\mathbf{v}_N$ then gives
$\|\xi_N-\hat\xi_N\|_1\le\lambda_N+|1-\lambda_N-\|\mathbf{v}_N\|_1|+\lambda_N\le3\lambda_N$. Uniform continuity of
$\mathrm{eff}_g$ and $\mathrm{eff}_{\tilde g}$ on the simplex transfers (a) and (b) when $\lambda_N\to0$. For a fixed $\lambda<1$ the
construction is a continuous map of $\hat\xi_N$, because $\|\mathbf{v}_N\|_1\ge1-\lambda>0$, and (b) gives the limit.
\end{proof}

Proposition~\ref{prop:plugin} concerns the design and its efficiency under the true field. It says nothing
about the realised mean squared error of the estimator on a data-chosen design, which
Proposition~\ref{prop:adaptive} covers only for a fixed design (Remark~\ref{rem:threshold}). Part (a) needs a field
estimate that is consistent for the field itself on the whole grid, for instance a correctly specified
parametric model of the cumulant field or a pilot spread over the grid. The two-stage procedure of
Section~\ref{subsec:kurt} estimates the field only at the three pilot doses and interpolates it linearly. The
plug-in of \eqref{eq:g3} is a continuous function of the sample cumulants, so its estimate converges to the
linear interpolant $\tilde g$ of the field through the pilot doses, not to the field; parts (b) and (c) then give
the limit of that procedure, which Section~\ref{subsec:kurt} evaluates.

\section{When does the design move? A degeneracy dichotomy}
\label{sec:dichotomy}

The transported equivalence theorem reduces $\PMM_S$-optimal design to ordinary optimal design in the gauged regressor $\bh(x)=\gS(x)^{-1/2}\bef(x)$. The structural question is then for which error laws the optimal allocation differs from the classical one built on $\bef$. For a fixed design the relevant condition is whether $\Mg(\xi)$ is proportional to $\bM(\xi)$; across all designs, a field constant on $\mathcal{X}$ guarantees the classical optimiser. Only a varying field can move the optimum, and a varying field is necessary for separation but not sufficient: whether the optimiser moves then depends on the field, the criterion and the support. The results below make this precise and then locate the boundary between the second-order ($\SLS$) theory and the higher-order ($\PMM_S$, $S\ge3$) reserve.

\begin{theorem}[Degeneracy dichotomy]
\label{thm:dichotomy}
Fix a degree $S\ge2$ and let $\xi$ range over approximate designs on $\mathcal{X}$ with support points $\{x_k\}$ and weights $\{w_k\}$.
\begin{enumerate}
\item[(a)] \emph{(Degenerate branch.)} If the reduction field is constant on the support, $\gS(x_k)\equiv g$ for all $k$ with $w_k>0$, then
\[
\Mg(\xi)=g^{-1}\,\bM(\xi).
\]
If $\gS\equiv g$ on the whole design region $\mathcal{X}$, then consequently, for every standard information criterion $\Psi$ (in particular $D$-, $A$-, and $c$-optimality), the $\PMM_S$-optimal design coincides with the classical $\OLS$-optimal design and does not depend on the value of $g$ nor on the error cumulants that determine it.
\item[(b)] \emph{(Separating branch.)} Let $\bef(x)=(1,x,x^2)^\top$ be the quadratic regression on $\mathcal{X}\subset\R$ of Section~\ref{sec:designinfo}. If $\xi$ has at most five distinct support points and $\gS$ is non-constant on its support, then $\Mg(\xi)$ is not a scalar multiple of $\bM(\xi)$, so the degenerate branch does not apply. Whether the $\PMM_S$-optimal design then differs from the $\OLS$-optimal design depends on the field and the criterion; Example~\ref{ex:kurt} below gives a numerical instance.
\end{enumerate}
In words: design separation requires a reduction field that is not constant on $\mathcal{X}$, and the classical homoscedastic-cumulant case is a degenerate boundary.
\end{theorem}

\begin{proof}
\emph{(a)} By the definition of the $\gS$-weighted information,
\[
\Mg(\xi)=\sum_{k} \frac{w_k}{\gS(x_k)}\,\bef(x_k)\bef(x_k)^\top .
\]
If $\gS(x_k)=g$ for every $k$ with $w_k>0$, the common factor $g^{-1}$ pulls out of the sum, giving $\Mg(\xi)=g^{-1}\sum_k w_k\,\bef(x_k)\bef(x_k)^\top=g^{-1}\bM(\xi)$.

It remains to show that, when $\gS\equiv g$ on all of $\mathcal{X}$, so that $\Mg(\xi)=g^{-1}\bM(\xi)$ for every design $\xi$ on $\mathcal{X}$, scaling the information matrix by the $\xi$-independent constant $g^{-1}>0$ leaves the optimal design unchanged. Each standard criterion is a function of the information matrix that is homogeneous of fixed degree under positive scaling. For $D$-optimality, $\det\!\big(g^{-1}\bM(\xi)\big)=g^{-p}\det\bM(\xi)$; for $A$-optimality, $\tr\!\big((g^{-1}\bM(\xi))^{-1}\big)=g\,\tr\bM(\xi)^{-1}$; for $c$-optimality, $c^\top\!\big(g^{-1}\bM(\xi)\big)^{-1}c=g\,c^\top\bM(\xi)^{-1}c$. More generally, every Kiefer $\Phi_q$ criterion is positively homogeneous in the information matrix. In each case the objective evaluated at $\Mg$ equals the objective evaluated at $\bM$ times a fixed power of $g$---a $\xi$-independent positive constant. A strictly monotone rescaling of an objective by a $\xi$-independent positive constant preserves its set of maximisers (respectively minimisers) over the design space. Hence
\[
\arg\Psi\big[\Mg(\cdot)\big]=\arg\Psi\big[\bM(\cdot)\big],
\quad\text{i.e.}\quad
\xi^{*}_{\PMM_S}=\xi^{*}_{\OLS}.
\]
Since $g$ survives only as the discarded common factor, the optimiser is independent of $g$ and therefore of the constant cumulants underlying it.

\emph{(b)} Suppose $\gS(x_k)$ is non-constant on the support, with $\gS(x_j)\neq\gS(x_l)$ for two points $x_j,x_l$ ($w_j,w_l>0$). First, $\Mg(\xi)$ is not a scalar multiple of $\bM(\xi)$. Indeed, for any $c>0$,
\[
\Mg(\xi)-c\,\bM(\xi)=\sum_k w_k\big(\gS(x_k)^{-1}-c\big)\,\bef(x_k)\bef(x_k)^\top .
\]
For the quadratic regression the entries of $\bef(x)\bef(x)^\top$ are the monomials $1,x,\dots,x^4$, and the matrix is determined by the vector $\mathbf{v}(x)=(1,x,x^2,x^3,x^4)^\top$. For $K\le5$ distinct points $x_1,\dots,x_K$ the vectors $\mathbf{v}(x_k)$ are linearly independent, because the first $K$ coordinates $(1,x_k,\dots,x_k^{K-1})$ form a $K\times K$ Vandermonde matrix, whose determinant $\prod_{j<l}(x_l-x_j)$ does not vanish on distinct nodes; hence the rank-one matrices $\{\bef(x_k)\bef(x_k)^\top\}$ are linearly independent. Hence the displayed sum vanishes only if $w_k(\gS(x_k)^{-1}-c)=0$ for every $k$, i.e.\ $\gS(x_k)=c^{-1}$ at every $x_k$ with $w_k>0$ --- contradicting non-constancy. Thus $\Mg\neq c\,\bM$ for all $c>0$, and the scalar-invariance argument of part (a) is unavailable. By the transported equivalence theorem (Theorem~\ref{thm:equiv}) the $\PMM_S$-optimal design is then the ordinary $\Psi$-optimal design in the gauged regressors $\bh(x)=\gS(x)^{-1/2}\bef(x)$, which need not coincide with the $\Psi$-optimal design in $\bef$.
\end{proof}

\begin{example}[A numerical instance of separation]
\label{ex:kurt}
This example is a computation, not part of the proof of Theorem~\ref{thm:dichotomy}. The $\OLS$ $A$-optimal design on $[0,1]$ is supported on $\{0,t^\ast,1\}$ with $t^\ast=0.4998$ (found by minimising the criterion over the interior node; $t=\tfrac12$ itself is not exactly optimal) and weights proportional to the column norms of $\mathbf{F}^{-1}$, $\mathbf{F}$ the $3\times3$ matrix of rows $\bef(x_k)^\top$, i.e.\ $(0.322,0.486,0.192)$; the $A$-equivalence condition for $\bM$ holds at every point of a grid of $2\times10^5$ points of $[0,1]$, which is numerical support for its optimality, not a proof. Non-optimality for $\Mg$ needs no grid: it suffices that the condition fails at one point, and it fails at $x=0$. For the closed-form \textsc{kurt} field of \appref{app:g3}, $g_3(x)$ with $a(x)=8-7x$, this design gives $\max_x\bh(x)^\top\Mg^{-2}\bh(x)=131.8$ (at $x=0$) against $\tr\Mg^{-1}=110.4$, so it violates the $A$-equivalence condition for $\Mg$ and is not $A$-optimal there. Hence, for $S=3$, the $A$-criterion and this field, the $\PMM_3$-optimal design differs from the $\OLS$-optimal one; Section~\ref{subsec:kurt} reports both designs.
\end{example}

\begin{remark}[Why part (b) is restricted to five support points]
\label{rem:five}
With six or more support points the matrices $\bef(x_k)\bef(x_k)^\top$ span at most the five-dimensional space of Hankel matrices in the moments $1,x,\dots,x^4$, so they are linearly dependent and a non-constant field can leave $\Mg(\xi)$ proportional to $\bM(\xi)$. For instance, on the six equally spaced points $\{0,0.2,\dots,1\}$ with equal weights there is a non-constant field with values in $(0,1]$ for which $\Mg(\xi)=4.78\,\bM(\xi)$. Proportionality on one design does not make the field constant on $\mathcal{X}$, so part (a) does not apply either; for such designs the comparison must be made through the criterion itself. The optimal designs of Section~\ref{sec:numerical} have three support points.
\end{remark}

One point needs care before the dichotomy is applied to the second-order channel. The
closed form \eqref{eq:g2} gives $g_2<1$ under skewness, which would seem to make the $S=2$
field non-constant and to force separation of the locally-adaptive $\PMM_2$ design by
Theorem~\ref{thm:dichotomy}(b). It does not. The reduction that enters $\Mg$ is the
realised \emph{per-stratum location} reduction, and at second order, with the scale estimated
within the stratum, that reduction is identically one.

\begin{lemma}[The second-order location reduction is unity]
\label{lem:loc2}
For the per-stratum degree-2 $\PMM$ (equivalently $\SLS$) location estimator with a scale
co-estimated in the stratum, the realised reduction field is $g_2(x)\equiv1$ for every error law
with finite fourth moment, symmetric or skewed. Equivalently, the second-order skewness reserve
quantified by \eqref{eq:g2} is not realised in a location parameter whose stratum carries its own
scale.
\end{lemma}

\begin{proof}
Within a single stratum the regression mean $\mu_k=\bef(x_k)^\top\btheta$ is constant in the
sampling index, and the degree-2 estimating system consists of the two moment conditions
$\E[y-\mu_k]=0$ and $\E[y^2-\mu_k^2-\sigma_k^2]=0$ in the two unknowns $(\mu_k,\sigma_k^2)$. The
system is exactly identified, so every non-singular combination of the two equations has the
same root, namely $\hat\mu_k=\bar y_k$ and $\hat\sigma_k^2$ the stratum variance; the location
estimator is the sample mean and $g_2(x_k)=1$. The skewness reserve \eqref{eq:g2} is transmitted
through the term $2\mu e$ of the regression-form residual $y^2-\mu^2-\sigma^2$
\citep[Section~3.4]{zabolotnii2026slspmm}, and here a free $\sigma_k^2$ absorbs that direction.
Under a varying mean and a common scale the same term carries the reserve to the slope
\citep[Corollary~1]{zabolotnii2026slspmm}. The calibration of
\S\ref{subsec:kurt}--\S\ref{subsec:skew} confirms $g_2(x)=1.000$ on both a symmetric and a strongly
skewed field.
\end{proof}

Lemma~\ref{lem:loc2} is a statement about the locally-adaptive class, in which every stratum
carries its own scale. The model of this paper has one common variance, and an estimator that
uses it is not bound by the lemma. The natural example is the conditional $\SLS$ of
\citet{wang2008second}, which weights the regression-form residual vector $\rho_i$ by
$\{\E[\rho_i\rho_i^\top\mid x_i]\}^{-1}$ and has one parameter $\sigma^2$. By the change of variables
$\rho_i=T_i\,m(e_i)$, $m(e)=(e,\,e^2-\sigma^2)^\top$, that underlies the $\SLS=\PMM_2$ identity
\citep[Theorem~1 and Proposition~1]{zabolotnii2026slspmm}, its estimating function for
$(\btheta,\sigma^2)$ at an observation with regressor $x$ is
\begin{equation}
\label{eq:csls-score}
\psi(e,x)=\Big(\bef(x)\,\big[F_2(x)^{-1}m(e)\big]_1,\ \big[F_2(x)^{-1}m(e)\big]_2\Big),
\end{equation}
with the conditional body
$F_2(x)=\bigl(\begin{smallmatrix}\sigma^2&\kappa_3(x)\\ \kappa_3(x)&\mu_4(x)-\sigma^4\end{smallmatrix}\bigr)$
in place of the constant one. Write
$F_2(x)^{-1}=\bigl(\begin{smallmatrix}a(x)&c(x)\\ c(x)&d(x)\end{smallmatrix}\bigr)$; in standardised
cumulants $a=1/\{\sigma^2 g_2(x)\}$ with $g_2(x)$ the pointwise value of \eqref{eq:g2},
$c=-\gamma_3(x)/\{\sigma^3\Delta(x)\}$, $d=1/\{\sigma^4\Delta(x)\}$, $\Delta=2+\gamma_4-\gamma_3^2$.

\begin{proposition}[Second-order designs under the global conditional $\SLS$]
\label{prop:csls}
Assume a common error variance $\sigma^2$, known conditional moments through order four (an oracle benchmark) with
$\Delta(x)>0$, and an intercept in $\bef$. For a design $\xi$ the conditional $\SLS$ estimator of
$\btheta$ is asymptotically normal with covariance $N^{-1}\,\mathbf{I}_2(\xi)^{-1}$, where
\begin{equation}
\label{eq:csls-info}
\mathbf{I}_2(\xi)=\int a\,\bef\bef^\top d\xi-\frac{\big(\int c\,\bef\,d\xi\big)\big(\int c\,\bef\,d\xi\big)^\top}{\int d\;d\xi}.
\end{equation}
\begin{enumerate}
\item[(a)] If the cumulant field is constant, then
$\mathbf{I}_2(\xi)^{-1}=\sigma^2\{g_2\,\bM(\xi)^{-1}+(1-g_2)\,\mathbf{e}_1\mathbf{e}_1^\top\}$ with
$\mathbf{e}_1$ the intercept coordinate, so the $D$-, $A$- and $c$-optimal designs for $\btheta$
coincide with the $\OLS$-optimal ones.
\item[(b)] If the errors are symmetric, $\gamma_3(x)\equiv0$, then $\mathbf{I}_2(\xi)=\sigma^{-2}\bM(\xi)$
whatever the kurtosis field, and every optimal design coincides with the $\OLS$-optimal one.
\item[(c)] If the skewness varies on the support, $\mathbf{I}_2(\xi)$ is in general not of either
form, and the conditional-$\SLS$-optimal design can move with the field; on the \textsc{skew} field of Section~\ref{subsec:skew} it does.
\end{enumerate}
\end{proposition}

\begin{proof}
The estimating function \eqref{eq:csls-score} has $\E[\psi\psi^\top\mid x]=-\E[\partial\psi\mid x]
=\mathbf{J}(x)$ with $\mathbf{J}(x)=\bigl(\begin{smallmatrix}a\,\bef\bef^\top&c\,\bef\\ c\,\bef^\top&d\end{smallmatrix}\bigr)$,
because $\E[\partial m/\partial\btheta^\top\mid x]=(-\bef,\mathbf{0})^\top$,
$\partial m/\partial\sigma^2=(0,-1)^\top$ and $\E[mm^\top\mid x]=F_2(x)$. The joint information of
$(\btheta,\sigma^2)$ is therefore $\int\mathbf{J}\,d\xi$, additive in the design, and the standard
sandwich argument gives the covariance of $\hat\btheta$ as the inverse of its Schur complement,
which is \eqref{eq:csls-info}. (a) With constant cumulants, $c^2/d=(g_2^{-1}-1)/\sigma^2$ and
$\int\bef\,d\xi=\bM(\xi)\mathbf{e}_1$ because the first regressor is one, so
$\sigma^2\mathbf{I}_2=g_2^{-1}\bM-(g_2^{-1}-1)\bM\mathbf{e}_1\mathbf{e}_1^\top\bM$. The
Sherman--Morrison formula, with $\mathbf{e}_1^\top\bM\mathbf{e}_1=\int d\xi=1$, gives the stated
inverse. Then $\det\mathbf{I}_2=\sigma^{-2p}g_2^{1-p}\det\bM$,
$\tr\mathbf{I}_2^{-1}=\sigma^2\{g_2\tr\bM^{-1}+1-g_2\}$ and
$\mathbf{c}^\top\mathbf{I}_2^{-1}\mathbf{c}=\sigma^2\{g_2\,\mathbf{c}^\top\bM^{-1}\mathbf{c}+(1-g_2)c_1^2\}$;
each is an increasing affine function of the corresponding $\OLS$ criterion, so the optimisers
agree. (b) Under symmetry $\kappa_3(x)\equiv0$, hence $c\equiv0$ and $a\equiv1/\sigma^2$: the
optimal combination at every $x$ is $F_2(x)^{-1}b=(1/\sigma^2,0)^\top$, the pointwise form of
\citet[Proposition~2]{zabolotnii2026slspmm}, and \eqref{eq:csls-info} reduces to $\sigma^{-2}\bM$.
(c) When $\gamma_3$ varies, $a(x)$ and $c(x)$ vary and \eqref{eq:csls-info} is a genuine
reweighting with a design-dependent rank-one correction; Section~\ref{subsec:skew} computes the
resulting separation for the \textsc{skew} field.
\end{proof}

The optimal-design problem for \eqref{eq:csls-info} reduces to a standard one. The estimator
$\hat\btheta$ is the coefficient block of an estimator whose joint information $\int\mathbf{J}\,d\xi$ is
additive. $A$-optimality for $\btheta$ is therefore an $L$-optimality problem, and $D$-optimality for
$\btheta$ is a $D_s$-optimality problem with $\sigma^2$ as nuisance. Both are convex and are computed by the
usual multiplicative algorithms. This is the approximate theory of \citet{bose2015optimal},
in which the $\SLS$ information is nonlinear in the design measure, with the conditional body
allowed to vary.

\begin{theorem}[Second-order blindness in the symmetric class versus the third-order reserve]
\label{thm:reserve}
\leavevmode
\begin{enumerate}
\item[(i)] \emph{In the symmetric class second order is design-blind.} Let the errors satisfy the standing
assumptions of Section~\ref{sec:designinfo} at every dose (common variance $\sigma^2$; for each degree $S$
considered, finite moments of order $2S$ and a law not concentrated on $S$ or fewer points) and be symmetric
at every dose, $\gamma_3(x)\equiv0$ on the whole design region $\mathcal{X}$, with an arbitrary kurtosis
field $\gamma_4(x)$. Then for every design $\xi$ on $\mathcal{X}$ the
locally-adaptive $\PMM_2$ has $\Mg(\xi)=\bM(\xi)$ and the global conditional $\SLS$ has
$\mathbf{I}_2(\xi)=\sigma^{-2}\bM(\xi)$, so the optimal design of either coincides with the
$\OLS$-optimal design under every criterion. Neither second-order design responds to a symmetric
kurtosis field.
\item[(ii)] \emph{Third order can separate.} Under the assumptions of (i), the degree-3 reduction $g_3(x)$
acts on the location even when the within-stratum mean is constant and the scale and shape are
co-estimated (Lemma~\ref{lem:g3}), and it depends on the even cumulants through $g_3(x)=1-\gamma_4(x)^2/\{6+9\gamma_4(x)+\gamma_6(x)\}$, the closed form \eqref{eq:g3}.
For the quadratic regression $\bef(x)=(1,x,x^2)^\top$, when $g_3(x)$ is non-constant on the support of a
design with at most five distinct support points, $\bM_{g_3}$ is not a scalar multiple
of $\bM$ and, by Theorem~\ref{thm:dichotomy}(b), the $\PMM_3$-optimal design can separate from the
$\OLS$-optimal one (for the Beta family of Section~\ref{subsec:kurt} it does, by the computation of Example~\ref{ex:kurt}). For the two second-order estimators of part (i), design separation within
the symmetric class can thus occur only at order $S\ge3$.
\end{enumerate}
\end{theorem}

\begin{proof}
\emph{(i)} For the locally-adaptive $\PMM_2$, Lemma~\ref{lem:loc2} gives $g_2(x)\equiv1$ and
Theorem~\ref{thm:dichotomy}(a) with $g\equiv1$ gives $\Mg=\bM$. For the conditional $\SLS$ this is
Proposition~\ref{prop:csls}(b).
\emph{(ii)} By (i) no second-order design separates in the symmetric class. For $S=3$,
Lemma~\ref{lem:g3} in \appref{app:g3} shows that, under symmetry, the per-stratum reduction with
co-estimated scale and shape is $g_3=1-\gamma_4^2/(6+9\gamma_4+\gamma_6)$, which is below one whenever
$\gamma_4\ne0$ (at a dose with $\gamma_4=0$ the degree-$3$ equation has no reserve and $g_3=1$). A varying kurtosis
field can therefore make $g_3(x)$ non-constant on the support (for the Beta family of
Section~\ref{subsec:kurt} it is strictly monotone in the shape, Table~\ref{tab:g3closed}); when it does,
Theorem~\ref{thm:dichotomy}(b) applies. Since by (i) no second-order design separates in the symmetric class,
separation there can arise only at order $S\ge3$.
\end{proof}

\begin{remark}[Degrees above three]
\label{rem:higherS}
Part (ii) is stated for $S=3$, the degree with a closed-form reduction and the one the numerical study
uses. It carries over to every odd $S\ge3$. Under symmetry Lemma~\ref{lem:gS} gives the per-stratum
reduction $\gS(x)$ of \eqref{eq:gS-odd} at the local moments, so whenever $\gS$ is non-constant on a design
with at most five support points, $\bM_{g_S}$ is not proportional to $\bM$ and is the information of the
stratify-and-pool $\PMM_S$. An even degree adds nothing under symmetry ($g_{2r}=g_{2r-1}$), and
$g_{S+2}\le\gS$ whenever the moment and support conditions also hold at degree $S+2$. A non-constant $g_3$ does not by itself make $\gS$ non-constant for $S>3$, since a higher
degree can equalise the reductions.
\end{remark}

\paragraph{Scope of the blindness claim}
Theorem~\ref{thm:reserve}(i) is a statement about the symmetric class. It does not say that
second-order designs are blind in general. Proposition~\ref{prop:csls}(c) shows the opposite for a
varying skewness field: the global conditional $\SLS$, which uses the common error variance, moves
its design, and Section~\ref{subsec:skew} measures by how much. Only the locally-adaptive $\PMM_2$,
whose strata each carry their own scale, stays blind there (Lemma~\ref{lem:loc2}). In the
cumulant-homoscedastic case, which is the i.i.d.\ setting of the $\SLS$ optimal-design line
\citep{gao2014optimal,bose2015optimal,he2018roptimality,he2024comparing,he2026dose},
Proposition~\ref{prop:csls}(a) covers the coefficient block of a linear model with an intercept
under the $D$-, $A$- and $c$-criteria. Designs in that literature that differ from $\OLS$ designs do
so outside these conditions, for instance in models nonlinear in the parameters or without an
intercept. The claim of this paper is narrower: a symmetric field, however
strongly its kurtosis varies, gives a second-order design nothing to respond to, and this is
the regime that the third-order reserve opens.

The numerical study below is consistent with these branches. The symmetric, kurtosis-heteroscedastic \textsc{kurt} regime produces a $\PMM_3$ design separation of $0.545$ against $0$ for $\PMM_2$ and for the conditional $\SLS$. In the asymmetric \textsc{skew} regime the locally-adaptive $\PMM_2$ does not separate and the conditional $\SLS$ does, as Theorems~\ref{thm:dichotomy} and \ref{thm:reserve} and Proposition~\ref{prop:csls} predict.

Table~\ref{tab:estimators} collects the three estimators that these results compare.

\begin{table}[t]
\centering
\caption{On a symmetric field the second-order estimators keep the $\OLS$ information; only the
stratify-and-pool $\PMM_S$ with $S\ge3$ and the symmetry imposed realises a reduction there. For each estimator: what
is fitted, the limit of $N\Cov(\hat\btheta)$, and whether its optimal design can depart from the $\OLS$ design on a
symmetric field with varying kurtosis (\textsc{kurt}) and on a field with varying skewness (\textsc{skew}).}
\label{tab:estimators}
{\footnotesize\setlength{\tabcolsep}{3pt}
\begin{tabular}{>{\raggedright\arraybackslash}p{0.17\textwidth}>{\raggedright\arraybackslash}p{0.22\textwidth}>{\raggedright\arraybackslash}p{0.14\textwidth}p{0.07\textwidth}p{0.07\textwidth}>{\raggedright\arraybackslash}p{0.14\textwidth}}
\toprule
Estimator & Fit & limit of $N\Cov(\hat\btheta)$ & \textsc{kurt} & \textsc{skew} & Result\\
\midrule
Stratify-and-pool $\PMM_2$ & degree-2 location fit per dose with its own scale; pooled by least squares
  & $\sigma^2\bM(\xi)^{-1}$ & no & no & Lemma~\ref{lem:loc2}\\
Global conditional $\SLS$ & one estimating equation for all doses with a common variance
  & $\mathbf{I}_2(\xi)^{-1}$ & no & yes & Prop.~\ref{prop:csls}\\
Stratify-and-pool $\PMM_S$, $S\ge3$ & degree-$S$ fit per dose with its own scale and shape, symmetry imposed;
  pooled by generalised least squares & $\sigma^2\Mg(\xi)^{-1}$ & yes & no$^{a}$ & Thm.~\ref{thm:reserve}, Lemma~\ref{lem:gS}\\
\bottomrule
\end{tabular}}
\par\smallskip
{\footnotesize $^{a}$With a free shape the stratum fit is the stratum mean and the information is that of $\OLS$
(Lemma~\ref{lem:gS}(a)); with the symmetry imposed on a skewed stratum the root is biased
(Proposition~\ref{prop:asym}).}
\end{table}

\paragraph{Imposed symmetry under a small asymmetry}
The third-order reserve of Theorem~\ref{thm:reserve}(ii) is realised only because each stratum imposes the
symmetry (Lemma~\ref{lem:gS}(b)). On a skewed stratum the odd-power equation is not centred and its root is not
consistent for the mean (end of \appref{app:gS}). The next result prices a small asymmetry. It takes the
skewness of order $n^{-1/2}$, the scale on which the bias and the reserve are of the same size.

\begin{proposition}[Degree-3 location under local asymmetry]
\label{prop:asym}
For each $n$ let $y_i=\mu+e_{n,i}$, $i=1,\dots,n$, with $e_{n,1},\dots,e_{n,n}$ i.i.d.\ from a law $P_n$ with
$\E e_{n,1}=0$. Suppose that $\sup_n\E|e_{n,1}|^{6+\delta}<\infty$ for some $\delta>0$, that the central moments of
$P_n$ of order up to six converge to those of a law $P$ that is symmetric, is not concentrated on three or fewer
points and has $\gamma_4\neq0$, and that $\sqrt n\,\gamma_3(P_n)\to h\in\R$. Let $\hat\mu$ be the plug-in estimator of
Lemma~\ref{lem:g3}(c), and let $m_2$, $\gamma_4$, $\gamma_6$ and $g_3$ be those of $P$, with $D=6+9\gamma_4+\gamma_6$. Then
\[
\sqrt n\,(\hat\mu-\mu)\xrightarrow{\,d\,}N\!\Big(-\sqrt{m_2}\,\frac{\gamma_4}{D}\,h,\ m_2\,g_3\Big).
\]
The second moment of this limit, relative to the variance $m_2$ of the limit of $\sqrt n(\bar y-\mu)$, is
\begin{equation}
\label{eq:asym-mse}
g_3+(1-g_3)\,\frac{h^2}{D},
\end{equation}
which is below one if and only if $h^2<D$. Under $P$, $D$ is the asymptotic variance of $\sqrt n$ times the sample
skewness. The stratum therefore keeps a gain over its mean if and only if its skewness is smaller in absolute value
than one asymptotic standard error of its own sample skewness.
\end{proposition}

The proof is in \appref{app:asym}. For the regression the bias passes through the pooling.

\begin{corollary}[Pooled bias]
\label{cor:asym}
In the setting of Proposition~\ref{prop:adaptive} with $S=3$ and the symmetry imposed in every stratum, let the
error law at $x_k$ satisfy the conditions of Proposition~\ref{prop:asym} along $n_k$ with
$\sqrt N\,\gamma_3(x_k)\to h_k$, limiting cumulants $\gamma_4(x_k)$ and $\gamma_6(x_k)$, and
$D_k=6+9\gamma_4(x_k)+\gamma_6(x_k)$. Then $\sqrt N(\hat\btheta-\btheta)\xrightarrow{d}N\big(\sigma\boldsymbol\beta,\ \sigma^2\Mg(\xi)^{-1}\big)$ with
\[
\boldsymbol\beta=-\Mg(\xi)^{-1}\sum_k\frac{w_k}{g_3(x_k)}\,\frac{\gamma_4(x_k)}{D_k}\,h_k\,\bef(x_k),
\]
so the limit of $N$ times the total mean squared error is $\sigma^2\{\tr\Mg(\xi)^{-1}+\|\boldsymbol\beta\|^2\}$. If
instead the skewness is fixed, $\gamma_3(x_k)\neq0$ at support points with $\gamma_4(x_k)\neq0$, the stratum roots there
are biased. Unless the vector of these stratum biases lies in the null space of $\mathbf{F}^\top\mathbf{W}$, with
$\mathbf{F}$ the matrix of rows $\bef(x_k)^\top$ and $\mathbf{W}=\mathrm{diag}\{w_k/g_3(x_k)\}$, the bias of $\hat\btheta$ does
not vanish and $N$ times its mean squared error diverges, whereas that of $\OLS$ stays bounded. For the quadratic
on three support points $\mathbf{F}$ is invertible, so any non-zero stratum bias suffices.
\end{corollary}

\begin{proof}
Along $n_k=Nw_k$ the stratum skewness satisfies $\sqrt{n_k}\,\gamma_3(x_k)\to\sqrt{w_k}\,h_k$, so by
Proposition~\ref{prop:asym}
\[
\sqrt N(\hat\mu_k-\mu_k)\xrightarrow{\,d\,}N\big(-\sigma\,(\gamma_4(x_k)/D_k)\,h_k,\ \sigma^2g_3(x_k)/w_k\big),
\]
independently over $k$. The plug-in reductions remain consistent for $g_3(x_k)$. The pooling of \appref{app:prop1} is linear with
coefficient matrix converging to $\Mg(\xi)^{-1}\mathbf{F}^\top\mathbf{W}$, which maps
the vector of stratum biases to $\sigma\boldsymbol\beta$ and the covariance to $\sigma^2\Mg(\xi)^{-1}$. For a fixed skewness the
population root $t$ of Lemma~\ref{lem:g3} solves $t^3+(3m_2-\kappa^\ast)t=m_3$ (\appref{app:asym}), so it differs from
$\mu_k$ whenever $m_3\neq0$. The limit of the pooled bias is $\Mg(\xi)^{-1}\mathbf{F}^\top\mathbf{W}$ times the vector of these
stratum biases, which is non-zero outside the null space of $\mathbf{F}^\top\mathbf{W}$, and then $N$ times the squared bias
diverges. The stratum mean used by $\OLS$ is unbiased.
\end{proof}

\begin{remark}[Finite-sample threshold]
\label{rem:threshold}
The reserve is an asymptotic property of the field $\gS$, and the locally-adaptive $\PMM_S$ realises
it only when each high-reserve stratum carries enough observations to estimate the cumulants
that control its local reduction. At degree $S=3$ the reduction couples to cumulants up to order
$\gamma_6$, and small per-stratum samples make the plug-in $\hat\gamma_6$ noisy. Estimating the
coefficients from sample moments does not change the first-order law of a degree-$S$ $\PMM$
estimator \citep[Proposition~3]{zabolotnii2026slspmm}; the cost is a finite-sample one, and it is
largest where the per-stratum samples are smallest \citep{zabolotnii2026slspmm,zabolotnii2026pmmfp}. We therefore
expect, but do not prove, a range of small $N$ in which the plug-in noise hides the design gain. For a
fixed design with fixed weights, Proposition~\ref{prop:adaptive} gives the analytic gain as the limit
when every $n_k\to\infty$; no uniform statement over data-chosen designs is made. The global-moment
$\PMM_S$, which pools cumulants across strata, cannot localise the per-stratum reduction, and its point
estimates showed no design gain at any $N$ we ran. The $N$-sweep below explores this range empirically.
\end{remark}

\section{Numerical study}
\label{sec:numerical}

This section checks the structural claims of Section~\ref{sec:dichotomy} numerically, through four questions. (Q1)~Does a symmetric kurtosis field leave the second-order designs at the $\OLS$ design, as Theorem~\ref{thm:reserve}(i) predicts? (Q2)~Does it move the third-order design, and by how much (Theorem~\ref{thm:reserve}(ii))? (Q3)~Does a varying skewness field, with the kurtosis tied to it, move the conditional-$\SLS$ design (Proposition~\ref{prop:csls})? (Q4)~Does the locally-adaptive $\PMM_3$ estimator of Proposition~\ref{prop:adaptive} realise the analytic gain at finite $N$, first with the field known and then with it estimated? (Q5)~How much asymmetry does that gain tolerate (Proposition~\ref{prop:asym})? Section~\ref{subsec:kurt} answers Q1, Q2, Q4 and Q5, and Section~\ref{subsec:skew} answers Q3. In brief, a symmetric kurtosis field leaves both second-order designs at the $\OLS$ design, and the third-order design moves, with a modest oracle gain of $4.7\%$. A varying skewness moves the conditional-$\SLS$ design. With the field known the gain is realised at $N\ge600$ in the separate run of Table~\ref{tab:twostage}; with the field estimated from a pilot it is unstable. The gain requires the absolute skewness of every stratum to stay below about one standard error of its sample skewness.

\subsection{Setup}
\label{subsec:setup}

The engine is R with the package EstemPMM (see Code availability): version~0.4.0 for the calibration and the Monte Carlo comparisons, version~0.5.0 for the two-stage study of Table~\ref{tab:twostage}. We use the quadratic dose--response model $y=\theta_0+\theta_1 x+\theta_2 x^2+\varepsilon$ with $\bef(x)=(1,x,x^2)^\top$, true parameters $\btheta=(2,\,1.5,\,-1)$, unit error variance $\sigma^2=1$, and design region $x\in\mathcal{X}=[0,1]$. Designs are optimised over the grid $\{0,0.02,\dots,1\}$ under $A$-optimality (total coefficient MSE): $\tr\,\Mg(\xi)^{-1}$ for the $\PMM$ designs and $\tr\,\mathbf{I}_2(\xi)^{-1}$ for the conditional $\SLS$, by multiplicative algorithms started from uniform weights on the grid and run until the largest change in a weight is below $10^{-11}$ ($10^{-12}$ for the conditional $\SLS$). For the $\OLS$ $A$-optimal design the largest excess of the $A$-sensitivity function over $\tr\bM^{-1}$ on the grid is then below $10^{-8}$. All Monte Carlo comparisons use $M=2500$ replications at a stated total sample size $N$. For the conditional $\SLS$ the update is $w_k\leftarrow w_k\{d_k/\sum_jw_jd_j\}^{1/2}$ with $d_k=\tr\{\mathbf{J}^{-1}\mathbf{L}\mathbf{J}^{-1}\mathbf{J}(x_k)\}$, $\mathbf{J}=\int\mathbf{J}(x)\,d\xi$ the joint information of Proposition~\ref{prop:csls} and $\mathbf{L}$ the selector of $\btheta$ (and its $D_s$ analogue). An approximate design is turned into stratum sizes by rounding each $Nw_k$ to the nearest integer. For the $\OLS$- and $\PMM_3$-optimal $A$-designs of the Monte Carlo comparisons the rounded sizes sum to $N$ at every $N$ we ran, and the recomputed $\PMM_3$ design also has an $A$-equivalence residual below $10^{-8}$. The two-stage study rounds by largest remainders. The pooling weights of the locally-adaptive estimator use the realised stratum sizes. Common random numbers couple the competing designs: each design is run from the same seed, so the error draws coincide wherever the two dose vectors coincide, which removes much of the sampling noise from the contrast.

\paragraph{Calibration of the reduction fields}
The degree-graded reduction fields $g_2(x)$ and $g_3(x)$ are not taken from closed forms. They are \emph{calibrated empirically} from EstemPMM's own per-stratum location fits, with $M_{\mathrm{cal}}=2500$ calibration replications, so the reported $\gS(x)$ are the reductions the estimator delivers rather than nominal ones. For each value of the shape parameter on a grid, each replication draws a sample of size $400$ from the error law and computes the sample mean and the intercept-only EstemPMM fits of degree $2$ and $3$ (scale co-estimated); $\gS$ is the ratio of the Monte Carlo variance of the degree-$S$ estimate to that of the sample mean, and the field is linearly interpolated between grid values. The calibration runs on its own random-number seeds, separate from those of the design comparison. The calibrated $g_2$ equals one to at least six decimal places at every grid value of both regimes, as Lemma~\ref{lem:loc2} requires: the intercept-only degree-2 fit returns the sample mean itself. The calibration and the main study ran on EstemPMM~0.4.0; we re-checked the identity directly in version~0.5.0, where the two differ by less than $2\times10^{-15}$. The $\PMM_2$ designs are therefore the designs for $g_2\equiv1$, and the $\PMM_2$ separation of order $10^{-15}$ reported below is floating-point rounding, not a design effect. As Monte Carlo estimates, the calibrated $\gS$ carry sampling error, of order a few thousandths near the no-reserve boundary of the \textsc{kurt} grid (e.g.\ $g_3=1.001$ at $a=8$). On the skewed \textsc{skew} grid the calibrated $g_3$ exceeds one at every grid value (from $1.007$ to $\approx1.12$). We read this as a finite-sample loss of the degree-$3$ fit with co-estimated scale on skewed errors rather than sampling noise. All seven values exceed one, six of them by $0.016$ or more, well above the few thousandths seen on the \textsc{kurt} grid; we did not confirm this reading with a larger calibration sample. The \textsc{skew} designs under test use $g_2$, not $g_3$. Values above the asymptotic ceiling $1$ are clipped to $1$ when used as design weights; on the \textsc{kurt} grid this affects only the value at $a=8$ ($1.001$). Because the calibration uses samples of size $400$, the calibrated field is a fixed-$n$ proxy for the asymptotic $\gS(x)$ that enters $\Mg$. The design analysis and the pooling weights treat it as a known (oracle) field. It is not a consistent estimate in the sense of Proposition~\ref{prop:adaptive}, so we gauge its effect on the design by replacing it with the closed form of \appref{app:g3}. The closed form gives the same $A$-optimal design at the grid resolution and an analytic gain of $1.048$ instead of $1.047$. We do not report Monte Carlo standard errors for the calibrated values; on the \textsc{kurt} grid they differ from the closed form by at most $0.03$ (Table~\ref{tab:g3closed}), which includes both the finite-sample and the simulation error.

We summarise the displacement of one design from another by the total-variation distance between their weight vectors on the common grid,
\[
\mathrm{sep}(\xi_1,\xi_2)=\tfrac12\sum_{x}\,\lvert w_1(x)-w_2(x)\rvert\in[0,1],
\]
which charges relocated design mass ($0$ = identical designs, $1$ = disjoint support). A support point that shifts to an adjacent grid node relocates its full weight, so $\mathrm{sep}$ registers a large value even for a geometrically small move. We therefore use $\mathrm{sep}$ only to show \emph{whether} a design moves. The efficiency gain, which unlike $\mathrm{sep}$ is comparable across criteria, measures \emph{how much} the move helps.

The error law is varied across two regimes. In the \textsc{kurt} regime the errors are symmetric and only the kurtosis varies with the regressor; this is the regime in which third-order design separation is predicted. The errors are centred Beta$(a(x),a(x))$ variates rescaled to unit variance with $a(x)=8-7x$, and the calibration grid is $a\in\{8,5,3.5,2.5,2,1.6,1.3,1\}$ (\S\ref{subsec:kurt}). In the \textsc{skew} regime the skewness is the varying cumulant, and the kurtosis follows it: the errors are centred Gamma variates of unit variance with $\gamma_3(x)=0.2+1.6x$ and $\gamma_4=1.5\gamma_3^2$, and the calibration grid is $\gamma_3\in\{0.2,0.5,0.8,1.1,1.4,1.6,1.8\}$ (\S\ref{subsec:skew}). There the locally-adaptive $\PMM_2$ is predicted not to separate (Lemma~\ref{lem:loc2}) and the global conditional $\SLS$ to separate (Proposition~\ref{prop:csls}(c)). The conditional-$\SLS$ designs are computed from the closed-form cumulants of the two error families; no calibration is needed, since \eqref{eq:csls-info} involves only $\gamma_3$ and $\gamma_4$. Their asymptotic covariance is checked by a Monte Carlo with the oracle conditional weights. Each regime is run both cumulant-heteroscedastically (the cumulant field varies over the support) and cumulant-homoscedastically (the field is held constant; the variance is one throughout); the homoscedastic run is the degeneracy control of Theorem~\ref{thm:dichotomy}.

\subsection{KURT regime: the third-order design separation}
\label{subsec:kurt}

The \textsc{kurt} errors are centred Beta$(a(x),a(x))$ variates rescaled to unit variance, with shape $a(x)=8-7x$: near-Gaussian at $x=0$ ($a=8$) and exactly uniform at $x=1$ ($a=1$). The family is symmetric throughout, so $\gamma_3(x)\equiv0$ and the platykurtic excess kurtosis $\gamma_4(x)$ is the only varying shape cumulant.

\begin{figure}[t]
\centering
\includegraphics[width=0.7\textwidth]{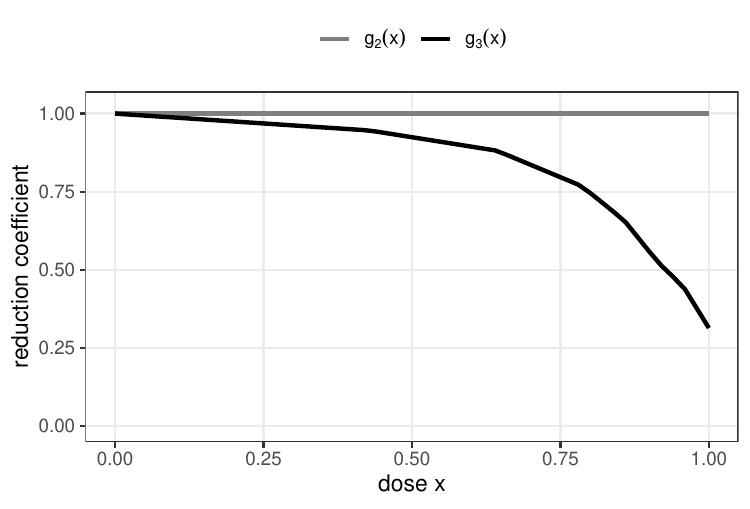}
\caption{The second-order reduction field is flat at one, while the third-order field falls across the dose range. Calibrated second- and third-order reduction fields $g_2(x)$ and $g_3(x)$ on the design region $x\in[0,1]$ for the \textsc{kurt} regime (symmetric Beta errors whose kurtosis varies with the dose), shape $a(x)=8-7x$; the varying $g_3$ is the reduction field induced by the cumulant heteroscedasticity, and it drives design separation.}
\label{fig:reserve_fields}
\end{figure}

\paragraph{Calibrated reduction fields}
Table~\ref{tab:calib_kurt} reports the calibrated reductions across the shape range. As Theorem~\ref{thm:reserve} requires in the symmetric class, the second-order reduction is degenerate, $g_2(x)\equiv1.000$ at every $x$, so the $\PMM_2$- and conditional-$\SLS$-optimal designs coincide with the $\OLS$ design here. The third-order field is non-constant. It falls monotonically from $g_3\approx1.00$ at $a=8$ to $g_3\approx0.31$ at $a=1$ as the error law moves from near-Gaussian to uniform. Down to $a=1$ the reduction stays a regular ($\sqrt{n}$-rate) one bounded away from $0$: the Beta law has all moments, and Lemma~\ref{lem:g3} of \appref{app:g3} gives the $\sqrt n$ limit with the closed-form value $g_3(1)=0.300$. The calibrated $0.313$ is a finite-sample ratio at $n=400$ and carries Monte Carlo error (\appref{app:g3}, Table~\ref{tab:g3closed}). Figure~\ref{fig:reserve_fields} plots the flat second-order field $g_2(x)$ and the descending third-order field $g_3(x)$ over the support. The flat $g_2$ predicts no second-order design movement, and the varying $g_3$ is what moves the third-order design in Table~\ref{tab:separation}.

\begin{table}[t]
\centering
\caption{For symmetric Beta$(a,a)$ errors, centred and rescaled to unit variance, the second-order reduction is one at every shape, while the third-order reduction falls from $1.00$ to $0.31$ as $a$ runs from $8$ (near-Gaussian) to $1$ (uniform). Entries are calibration estimates: the ratio of the Monte Carlo variance of the degree-$S$ polynomial-maximization location estimate ($g_2$ column: degree-2 fit; $g_3$ column: degree-3 fit), computed with the R package EstemPMM, to that of the sample mean, from $M_{\mathrm{cal}}=2500$ samples of size $400$ with unit variance (\textsc{kurt} regime: symmetric errors whose kurtosis varies with the dose).}
\label{tab:calib_kurt}
\begin{tabular}{rrr}
\toprule
Beta shape $a$ & $g_2(a)$ & $g_3(a)$ \\
\midrule
8.0 & 1.000 & 1.001 \\
5.0 & 1.000 & 0.946 \\
3.5 & 1.000 & 0.881 \\
2.5 & 1.000 & 0.768 \\
2.0 & 1.000 & 0.659 \\
1.6 & 1.000 & 0.524 \\
1.3 & 1.000 & 0.446 \\
1.0 & 1.000 & 0.313 \\
\bottomrule
\end{tabular}

\end{table}

\begin{figure}[t]
\centering
\includegraphics[width=0.7\textwidth]{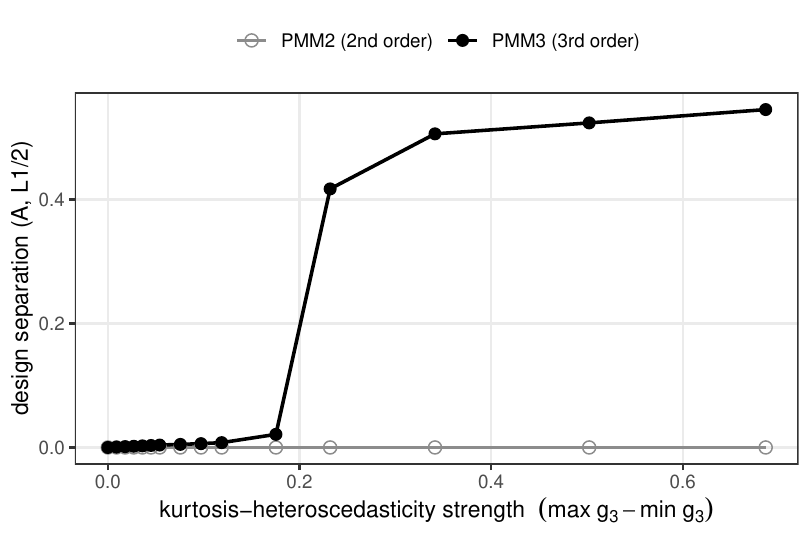}
\caption{Separation of the locally-adaptive third-order ($\PMM_3$) design grows with the strength of the kurtosis heteroscedasticity, while that of the second-order ($\PMM_2$) design stays at zero. Strength is the range $\max g_3-\min g_3$ of the calibrated third-order reduction field over $[0,1]$ for the shape fields $a(x)=\max(8-sx,1)$, $s=0,0.5,\dots,7$ ($s=0$ is homoscedastic, $s=7$ is the \textsc{kurt} field); separation is the total-variation distance from the $A$-optimal (minimum total coefficient variance) design of ordinary least squares ($\OLS$).}
\label{fig:separation_vs_strength}
\end{figure}

\paragraph{Design separation}
Table~\ref{tab:separation} summarises the separation between each $\PMM_S$-optimal design and the $\OLS$-optimal design across regimes, and Figure~\ref{fig:separation_vs_strength} traces it as a function of heteroscedasticity strength. In the heteroscedastic \textsc{kurt} case the second-order design separation is $0$ to machine precision ($2.4\times10^{-15}$) for the locally-adaptive $\PMM_2$, and exactly $0$ for the global conditional $\SLS$ (Table~\ref{tab:csls}): both second-order designs coincide with the $\OLS$-optimal design, as Theorem~\ref{thm:reserve}(i) predicts. The third-order design separation is $0.545$. In Figure~\ref{fig:separation_vs_strength} the third-order separation stays at or below $0.021$ while the strength is at most $0.18$, and lies between $0.42$ and $0.55$ for strengths from $0.23$ to $0.69$. The jump reflects the total-variation metric, which counts the full weight of a support point that moves to an adjacent grid node (\S\ref{subsec:setup}); the second-order separation is zero at every strength. What moves the design is the variation of $g_3$, not the degree of the estimator: at strength zero the third-order design does not move either. In the homoscedastic \textsc{kurt} control, where $g_3$ is held constant on the support, the third-order separation also returns to $0.000$, as the degeneracy dichotomy of Theorem~\ref{thm:dichotomy} requires: with a constant field, $\Mg=g^{-1}\bM$ is a scalar multiple of $\bM$ and the $\PMM_3$-optimal design coincides with the $\OLS$-optimal one.

\begin{table}[t]
\centering
\caption{Only the third-order design in the heteroscedastic \textsc{kurt} regime separates from the ordinary-least-squares ($\OLS$) design: separation $0.545$, analytic gain ratio $1.047$ ($+4.7\%$). Rows: separation of the locally-adaptive polynomial-maximization ($\PMM_S$) optimal designs from the $\OLS$-optimal design, by regime and cumulant field; the locally-adaptive second-order ($\PMM_2$) separation is $0$ everywhere (Lemma~\ref{lem:loc2}). The final column is the deterministic analytic $A$-efficiency gain ($A$: total coefficient variance, the trace criterion) of the $\PMM_S$-optimal over the $\OLS$-optimal design under the $g_S$-weighted criterion, with $S=3$ in the \textsc{kurt} rows and $S=2$ in the \textsc{skew} rows. Separation is the total-variation distance between each listed design and the corresponding $\OLS$-optimal design (\S\ref{subsec:setup}); ``---'' marks the \textsc{skew} rows, where $\PMM_3$ is not the estimator under test. \textsc{kurt}: symmetric errors, dose-varying kurtosis; \textsc{skew}: asymmetric errors, dose-varying skewness. The global conditional second-order least squares estimator ($\SLS$) is reported separately in Table~\ref{tab:csls}.}
\label{tab:separation}
{\small\setlength{\tabcolsep}{4pt}\begin{tabular}{llccc}
\toprule
Regime & Cumulant field & PMM$_2$-sep & PMM$_3$-sep & analytic $A$-gain \\
\midrule
KURT (symmetric) & heteroscedastic & 0.000 & 0.545 & 1.047 \\
KURT (symmetric) & homoscedastic & 0.000 & 0.000 & 1.000 \\
SKEW (asymmetric) & heteroscedastic & 0.000 & --- & 1.000 \\
SKEW (asymmetric) & homoscedastic & 0.000 & --- & 1.000 \\
\bottomrule
\end{tabular}
}
\end{table}

\begin{figure}[t]
\centering
\includegraphics[width=0.7\textwidth]{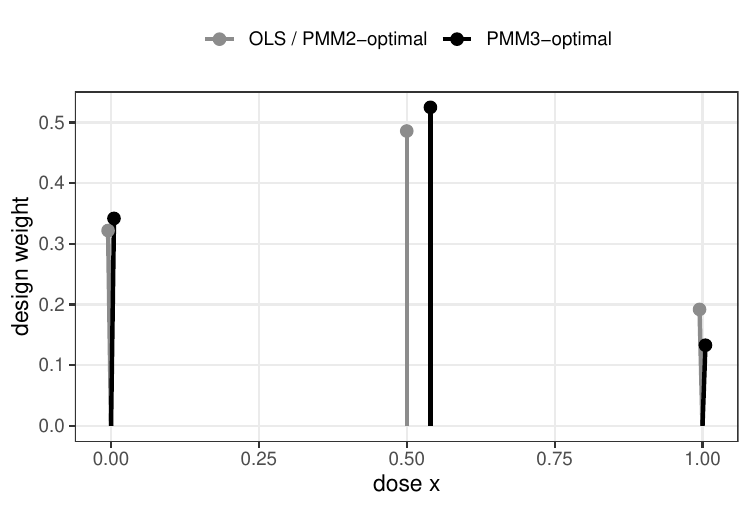}
\caption{The third-order design moves the interior dose from $0.50$ to $0.54$ and shifts weight off the top dose. $A$-optimal (minimum total coefficient variance) designs, written as support point\,:\,weight. Ordinary least squares and second-order polynomial maximization ($\OLS/\PMM_2$, identical designs): $\{0\!:\!0.32,\,0.50\!:\!0.49,\,1\!:\!0.19\}$; third-order polynomial maximization ($\PMM_3$): $\{0\!:\!0.34,\,0.54\!:\!0.53,\,1\!:\!0.13\}$. \textsc{kurt} regime: symmetric errors whose kurtosis varies with the dose.}
\label{fig:reserve_designs}
\end{figure}

\paragraph{The two optimal designs}
The displacement is reported in Figure~\ref{fig:reserve_designs}. The $\OLS/\PMM_2$-optimal $A$-design places mass at the three points
$\{0\!:\!0.32,\ 0.50\!:\!0.49,\ 1\!:\!0.19\}$ (support point : weight). The $\PMM_3$-optimal design responds to the $g_3$-weighting by shifting the interior support point from $0.50$ to $0.54$ and moving allocation off the top dose, which gives $\{0\!:\!0.34,\ 0.54\!:\!0.53,\ 1\!:\!0.13\}$ (weights summing to one). In this computation the variation of $g_3$ across the support drives the whole displacement. In general a non-constant field can redistribute design mass under a given criterion but need not (Theorem~\ref{thm:dichotomy} makes the variation necessary, not sufficient), while a flat field cannot.

\paragraph{Analytic efficiency gain}
Under the $g_3$-weighted ($A$) criterion the $\PMM_3$-optimal design improves on the $\OLS$-optimal design by $+4.7\%$ (efficiency ratio $1.047$). This figure involves no simulation beyond the calibration of the field, but it inherits the calibration error of the field. With the closed-form asymptotic field of \appref{app:g3} in place of the calibrated one the $A$-optimal design is unchanged at the grid resolution and the gain is $1.048$. Both gains use only the field and the two designs, and $1.047$ is the analytic target that the finite-sample realisation approaches. The gain is \emph{locally optimal} in the sense that both the design and the combining weights of the estimator treat the reduction field as known; Section~\ref{sec:discussion} discusses estimating the field.

\paragraph{Criterion comparison ($A$ versus $D$)}
Table~\ref{tab:criterion} repeats the heteroscedastic \textsc{kurt} computation under the $D$-criterion (generalised variance) alongside the $A$-criterion. The qualitative dichotomy does not depend on the criterion: under both criteria the second-order ($\PMM_2/\SLS$) separation is $0$ and only the third-order design separates. The \emph{magnitude} does depend on it. Under $D$ the gain is only $+0.1\%$, an order of magnitude below the $A$-gain. The $D$-optimal design for the quadratic keeps the three support weights at $\tfrac13$ each and can respond to the reserve field only by displacing the interior support point (from $0.50$ to $0.52$). The $A$-criterion can also redistribute the \emph{weights} and so obtains the larger $+4.7\%$. The $D$-separation metric registers $0.333$, the full interior weight relocating, but the efficiency it buys is negligible; separation must be read through the criterion. We therefore report $A$ as the headline. The $D$ result confirms the structural claim and shows how flat the generalised-variance criterion is here.

\begin{table}[t]
\centering
\caption{The third-order design gains $4.7\%$ under $A$-optimality (total coefficient variance) but only $0.1\%$ under $D$-optimality (generalised variance); gains are efficiency ratios relative to the ordinary-least-squares-optimal ($\OLS$) design for the same criterion (\textsc{kurt}-heteroscedastic regime: symmetric errors, dose-varying kurtosis). Second-order polynomial-maximization ($\PMM_2$) separation is $0$ under both criteria; the third-order design separates under both. The magnitude is criterion-specific: the $D$-gain ($+0.1\%$) is far below the $A$-gain ($+4.7\%$) because $D$-optimality fixes the support weights and can only nudge the interior support point.}
\label{tab:criterion}
{\footnotesize\setlength{\tabcolsep}{4pt}\begin{tabular}{lccc}
\toprule
Criterion & PMM$_2$ separation & PMM$_3$ separation & analytic PMM$_3$ gain \\
\midrule
$A$ (total coef.\ MSE) & 0.000 & 0.545 & 1.047 \\
$D$ (gen.\ variance) & 0.000 & 0.333 & 1.001 \\
\bottomrule
\end{tabular}
}
\end{table}

\begin{figure}[t]
\centering
\includegraphics[width=0.7\textwidth]{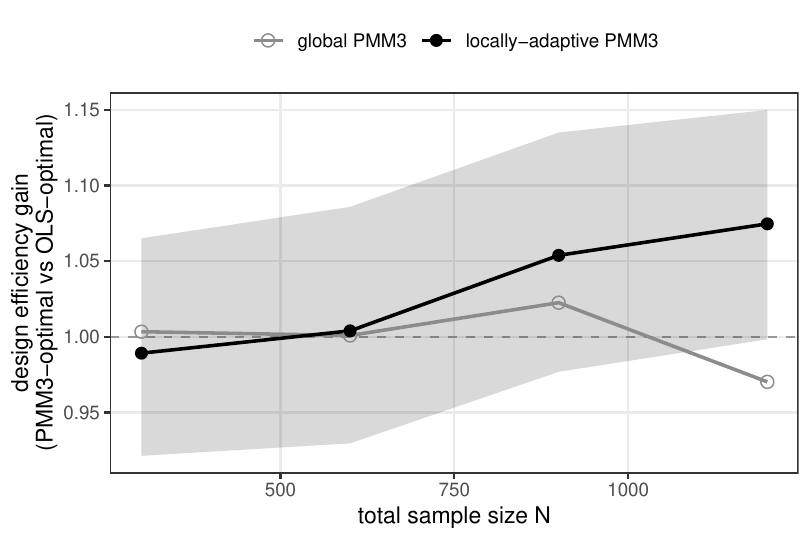}
\caption{Point estimates of the gain of the locally-adaptive third-order polynomial-maximization estimator ($\PMM_3$) are higher at $N\ge900$ than at $N\le600$, but the intervals are wide and all include one; the global-moment $\PMM_3$ point estimates stay near one. Realised ratio of total mean squared error, ordinary-least-squares-optimal design over $\PMM_3$-optimal design, against total sample size $N$; error bars are $95\%$ bootstrap intervals for the locally-adaptive estimator, all of which cover one.}
\label{fig:efficiency_vs_n}
\end{figure}

\paragraph{Finite-sample realisation across sample sizes}
Table~\ref{tab:nsweep} reports the realised total-MSE gain of the $\PMM_3$-optimal over the $\OLS$-optimal design as the total sample size $N$ grows, for both the locally-adaptive $\PMM_3$ of Proposition~\ref{prop:adaptive} and a global-moment $\PMM_3$ that pools cumulants across strata; Figure~\ref{fig:efficiency_vs_n} plots the same. The locally-adaptive estimator realises a gain of
$0.989\,[0.921,1.065]$ at $N=300$,
$1.004\,[0.930,1.086]$ at $N=600$,
$1.054\,[0.977,1.135]$ at $N=900$, and
$1.075\,[0.998,1.150]$ at $N=1200$ (point estimate with $95\%$ confidence interval). The point estimates are near one at $N\le600$ and lie at or slightly above the analytic $1.047$ at $N=900$ and $1200$ ($1.054$ and $1.075$). We read this descriptively: four sample sizes with overlapping intervals do not establish a trend or a threshold. It agrees with the behaviour anticipated in Remark~\ref{rem:threshold}, in which the controlling cumulants (degree $3$ couples to $\gamma_6$) are estimated noisily in small strata; convergence to $1.047$ is an $N\to\infty$ statement. We read these intervals conservatively. Every one of them, including the interval at $N=1200$ ($[0.998,1.150]$), still covers unity, so no single sample size establishes the gain on its own. This sweep is exploratory. A separate run of the same oracle comparison, made for the two-stage study below with other seeds and EstemPMM~0.5.0, gives gains of $1.035$ to $1.061$ at $N=600$ to $2400$ with every interval above one (Table~\ref{tab:twostage}). For the global-moment $\PMM_3$ we computed point estimates only, $1.003,\ 1.001,\ 1.023$ and $0.970$. Without intervals they are inconclusive and serve as no evidence for a contrast between the two estimators; they are merely compatible with the reserve being a \emph{per-stratum} (locally-adaptive) effect.

\begin{table}[t]
\centering
\caption{The point estimate of the locally-adaptive third-order polynomial-maximization ($\PMM_3$) gain is near one at $N\le600$ and near or above the analytic $1.047$ at $N\ge900$, while the global-moment estimator stays near one; no interval for the locally-adaptive estimator excludes one (none was computed for the global one). In each column the entry is the total mean squared error (sum over the three coefficients) of the ordinary-least-squares-optimal design divided by that of the $\PMM_3$-optimal design, both fitted with the estimator named in the column ($1.05$ means $5\%$ lower total MSE on the $\PMM_3$ design). The locally-adaptive estimator pools per-dose $\PMM_3$ location fits with weights $n_k/g_3(x_k)$; the global-moment estimator is one $\PMM_3$ regression fit with cumulants pooled over all doses. $95\%$ bootstrap confidence intervals for the locally-adaptive estimator only; $M=2500$ Monte Carlo replications per $N$, common random numbers (\textsc{kurt} regime).}
\label{tab:nsweep}
\begin{tabular}{rcc}
\toprule
Total $N$ & global PMM$_3$ & locally-adaptive PMM$_3$ (95\% CI) \\
\midrule
300 & 1.003 & 0.989 [0.921, 1.065] \\
600 & 1.001 & 1.004 [0.930, 1.086] \\
900 & 1.023 & 1.054 [0.977, 1.135] \\
1200 & 0.970 & 1.075 [0.998, 1.150] \\
\bottomrule
\end{tabular}

\end{table}

\paragraph{Degeneracy control}
As a direct check of Theorem~\ref{thm:dichotomy}, the homoscedastic \textsc{kurt} control (constant kurtosis field) yields a locally-adaptive realised gain of exactly $1.000$. Both arms use the same estimator, the locally-adaptive $\PMM_3$; the two optimal designs coincide, and common random numbers then make the two arms identical replication by replication, so the ratio is one by construction. This is a check of the design computation, not evidence about the estimator: without cumulant heteroscedasticity the $\PMM_3$-optimal design reproduces the $\OLS$-optimal one, and there is nothing to separate.

\paragraph{An estimated field: a two-stage design}
Everything above treats the field as known. Table~\ref{tab:twostage} reports a feasible version. A pilot of $N/3$ observations is placed on the $\OLS$-optimal design. In each pilot stratum $g_3$ is estimated by the plug-in of \eqref{eq:g3} (EstemPMM's own formula, clipped to $[0.05,1]$) and interpolated linearly in $x$ between the pilot doses. The $A$-optimal design for this estimated field is computed, and the remaining $2N/3$ observations fill the gap between the pilot allocation and $N$ times that design: each grid point receives a share of the second stage proportional to its positive deficit, rounded by largest remainders so that the counts are non-negative integers summing to $N$. The final estimate is the locally-adaptive $\PMM_3$ on all $N$ observations, pooled with plug-in weights. The comparators use the same $N$ and the same estimator: the one-stage $\OLS$-optimal design and the oracle one-stage $\PMM_3$-optimal design. This study is a separate run ($M=2500$ replications per $N$, its own seeds, EstemPMM~0.5.0, oracle field as calibrated above), so its oracle column replicates rather than repeats Table~\ref{tab:nsweep}. In it the oracle gain is $1.035$ to $1.061$, and all four intervals exclude one, including $[1.031,1.093]$ at $N=2400$. The two-stage design does not reliably realise the gain at these sample sizes. Its realised gain is $0.889$ $[0.824,0.959]$ at $N=600$, a loss, then $1.080$, $0.965$ and $1.051$ $[0.979,1.134]$ at $N=900$, $1200$ and $2400$, without a monotone trend. Two diagnostics locate the loss. Pooling the same two-stage data with oracle instead of estimated weights changes the realised gain by less than $0.007$ at every $N$ (Table~\ref{tab:twostage}); we found no material contribution of the pooling weights, although we did not compute an interval for this paired difference. The $A$-efficiency ratio of the design the two-stage procedure actually builds, evaluated under the true field, averages only $1.014$ to $1.020$. Most of this shortfall is structural and persists as $N\to\infty$ (Proposition~\ref{prop:plugin}(b),(c)). All pilot observations stay in the final design, and the field enters only through its values at the three pilot doses, interpolated linearly. With these three values known exactly, the $A$-optimal design for the interpolated field is $\{0\!:\!0.33,\ 0.58\!:\!0.52,\ 1\!:\!0.15\}$, whose $A$-efficiency gain under the calibrated field is $1.035$ instead of $1.047$, and the combined design with the one-third pilot reaches $1.023$. The designs actually built, $1.014$ to $1.020$, lie within one percentage point of that limit. The remaining gap is consistent with the noise of the pilot estimate of the $\gamma_6$-dependent field, which we did not measure. The interior mass is also split between the pilot dose $0.5$ and the new interior dose (on average $4.2$ to $4.3$ strata instead of three), so each interior stratum is smaller than in the one-stage design; this may add a finite-sample cost of the degree-$3$ fit, which we did not isolate. With a one-third pilot the design the feasible procedure builds therefore captures $29\%$ to $43\%$ of the oracle design gain at $N=600$ to $2400$ (design-only column against $1.047$), against $48\%$ in its limit, and its realised gain scatters around that value with wide intervals.

\begin{table}[t]
\centering
\caption{With the field known the $\PMM_3$ design gains $3.5$--$6.1\%$; with the field estimated from a one-third pilot the realised gain is unstable and the design itself is worth only $1.4$--$2.0\%$. Each realised entry is the total mean squared error of the one-stage ordinary-least-squares-optimal design divided by that of the procedure in the column, so values below one are losses; the definition applies to the oracle and to both two-stage columns ($95\%$ paired-bootstrap intervals for these realised columns only; $M=2500$ per $N$; locally-adaptive $\PMM_3$ in every arm). Oracle one-stage: the $\PMM_3$-optimal design for the calibrated field. Two-stage: pilot of $N/3$ on the $\OLS$ design, field estimated per pilot stratum, $A$-optimal augmentation; pooled with plug-in or with oracle weights. Design-only gain: mean $A$-efficiency ratio, under the calibrated field, of the allocation the two-stage procedure builds at that $N$. The oracle design has $1.047$. This column averages the analytic efficiency of each replication's design and carries no interval.}
\label{tab:twostage}
{\footnotesize\setlength{\tabcolsep}{3pt}\begin{tabular}{rcccc}
\toprule
 & oracle & \multicolumn{2}{c}{two-stage design} & design-only \\
Total $N$ & one-stage & plug-in weights & oracle weights & gain \\
\midrule
600 & 1.035 [1.005, 1.066] & 0.889 [0.824, 0.959] & 0.895 [0.830, 0.965] & 1.014 \\
900 & 1.059 [1.027, 1.091] & 1.080 [1.002, 1.167] & 1.086 [1.006, 1.173] & 1.017 \\
1200 & 1.057 [1.026, 1.089] & 0.965 [0.896, 1.039] & 0.966 [0.897, 1.039] & 1.018 \\
2400 & 1.061 [1.031, 1.093] & 1.051 [0.979, 1.134] & 1.054 [0.982, 1.138] & 1.020 \\
\bottomrule
\end{tabular}
}
\end{table}

\paragraph{Misspecified symmetry}
Proposition~\ref{prop:asym} was checked by Monte Carlo ($20\,000$ replications) on skewed $\mathrm{Beta}(a,a+d)$
errors with the estimator of Lemma~\ref{lem:g3}(c). For the uniform base law ($a=1$, $g_3=0.300$, $D=2.06$) and
$d=0.05$ the skewness is $0.042$. The bias of the population root, $0.024$ standard deviations, matches the
simulated bias ($0.023$ at $n=400$, $0.024$ at $n=2000$). The mean squared error relative to the stratum mean is
$0.545$ in theory and $0.547$ simulated at $n=400$. At $n=2000$ the same skewness is $1.3$ standard errors of the
sample skewness, and the ratio becomes a loss, $1.49$ in theory and $1.47$ simulated. For $a=2$ and $d=0.15$
(skewness $0.053$) the agreement is of the same order ($0.84$ against $0.85$ at $n=400$, $1.54$ against $1.55$ at
$n=2000$). The standard-error reading of $D$ was confirmed too: $n$ times the simulated variance of the sample
skewness at $n=2000$ is $2.05$ and $2.10$ against $D=2.06$ and $2.10$.

For the design, Corollary~\ref{cor:asym} was evaluated at the \textsc{kurt} designs with the closed-form field and a
skewness $\gamma$ constant across doses. At $\gamma=0$ the $\PMM_3$ design with the stratify-and-pool $\PMM_3$ has a
total $A$-criterion $1.286$ times smaller than the $\OLS$ design with $\OLS$; this combines the design gain $1.048$
with the gain of the estimator. To first order in $\gamma$ the advantage vanishes at $\sqrt N|\gamma|=7.7$, that
is $|\gamma|=0.22$ at $N=1200$ and $0.16$ at $N=2400$. The design gain alone, both designs fitted by the
stratify-and-pool $\PMM_3$, vanishes at $\sqrt N|\gamma|=10.3$ ($0.30$ and $0.21$). The pooled criterion tolerates more than a
single stratum, whose break-even skewness $\sqrt{D/n_k}$ is $0.07$ to $0.10$ at $n_k=400$, because its variance part
is dominated by the slope and curvature coefficients. Every tolerance shrinks like $N^{-1/2}$, so a fixed skewness
turns the advantage into a loss at large $N$. The numbers are produced by a script in the code supplement.

\subsection{SKEW regime: what the second-order channel can read}
\label{subsec:skew}

The \textsc{skew} regime varies the skewness, with the kurtosis tied to it. The errors are centred Gamma variates of unit variance with a skewness field $\gamma_3(x)=0.2+1.6x$, ranging from mild to strong asymmetry across the dose interval, and excess kurtosis $\gamma_4=1.5\,\gamma_3^2$; the homoscedastic control holds $\gamma_3\equiv1$. The kurtosis therefore varies too, and it enters $\mathbf{I}_2$ through $\Delta(x)$; but by Proposition~\ref{prop:csls}(b) a kurtosis field without skewness leaves the design where it is, so a non-zero skewness is what opens the second-order channel. Because $\gamma_4$ is tied to $\gamma_3$ in this family, the experiment does not apportion the size of the resulting move between the two cumulants.

\paragraph{Locally-adaptive $\PMM_2$}
The regime is a direct empirical test of Lemma~\ref{lem:loc2}. The calibrated second-order location reduction equals one to at least six decimal places across the entire field (Lemma~\ref{lem:loc2} makes it exact), although the skewness runs from $0.2$ to $1.8$. The calibration thus measures the location-blindness that the lemma asserts. Accordingly the locally-adaptive $\PMM_2$-optimal design coincides with the $\OLS$-optimal one (separation $0$, rows~3--4 of Table~\ref{tab:separation}), and the Monte Carlo design gain is exactly $1.000$ for both the global and the locally-adaptive estimators ($N=600$, $M=2500$). The two designs are identical, and common random numbers make the two arms coincide replication by replication.

\paragraph{Global conditional $\SLS$}
The conditional $\SLS$ of Proposition~\ref{prop:csls}, which uses the common error variance, does read the skewness field (Table~\ref{tab:csls}). Its $A$-optimal design on the heteroscedastic \textsc{skew} field is $\{0\!:\!0.35,\ 0.54\!:\!0.47,\ 1\!:\!0.18\}$ against the $\OLS$ design $\{0\!:\!0.32,\ 0.50\!:\!0.49,\ 1\!:\!0.19\}$: the interior dose moves to $0.54$ and weight shifts towards the low, least skewed dose. The separation is $0.501$ and the analytic $A$-efficiency gain over the $\OLS$-optimal design, for the same estimator, is $+2.5\%$ ($1.025$); under $D$-optimality the separation is $0.357$ and the gain $+0.6\%$ ($1.006$). In the homoscedastic control both separations are $0.000$, as Proposition~\ref{prop:csls}(a) requires, and on the symmetric \textsc{kurt} fields they are $0.000$ as well, as Theorem~\ref{thm:reserve}(i) requires. A Monte Carlo with the oracle conditional weights ($N=2000$, $M=40\,000$ replications, common random numbers) confirms the covariance \eqref{eq:csls-info}: the ratio of simulated to asymptotic total variance is $1.003$ on the $\OLS$-optimal design and $0.998$ on the conditional-$\SLS$-optimal design, and the realised total-MSE gain is $1.031$ ($95\%$ bootstrap interval $[1.012,1.051]$).

\begin{table}[t]
\centering
\caption{Only on the heteroscedastic \textsc{skew} field does the conditional second-order least squares ($\SLS$) design depart from the $\OLS$ design: separation $0.501$ ($A$) and $0.357$ ($D$), gain ratios $1.025$ and $1.006$. Global conditional $\SLS$ (Proposition~\ref{prop:csls}): separation (total-variation distance) of its $A$-optimal (minimum total coefficient variance) and $D$-optimal (minimum generalised variance) designs from the corresponding designs of ordinary least squares ($\OLS$), and the analytic efficiency gain of each over the $\OLS$-optimal design for the same estimator. Closed-form cumulant fields of the two error families (\textsc{kurt}: symmetric Beta errors, dose-varying kurtosis; \textsc{skew}: Gamma errors, dose-varying skewness); deterministic. The second-order design moves only on the varying skewness field; on the symmetric \textsc{kurt} fields and on the constant \textsc{skew} field it coincides with the $\OLS$ design.}
\label{tab:csls}
{\small\setlength{\tabcolsep}{4pt}\begin{tabular}{llcccc}
\toprule
Regime & Cumulant field & $A$-sep & $A$-gain & $D$-sep & $D$-gain \\
\midrule
KURT (symmetric) & heteroscedastic & 0.000 & 1.000 & 0.000 & 1.000 \\
KURT (symmetric) & homoscedastic & 0.000 & 1.000 & 0.000 & 1.000 \\
SKEW (asymmetric) & heteroscedastic & 0.501 & 1.025 & 0.357 & 1.006 \\
SKEW (asymmetric) & homoscedastic & 0.000 & 1.000 & 0.000 & 1.000 \\
\bottomrule
\end{tabular}
}
\end{table}

The \textsc{skew} regime therefore distinguishes the two second-order estimators rather than the two orders. A second-order design can respond to cumulant heteroscedasticity if the estimator uses the common variance and the heteroscedasticity is in the skewness. Neither the locally-adaptive $\PMM_2$ nor the conditional $\SLS$ (with a common variance and an intercept) can respond to a symmetric field, where the design movement of \S\ref{subsec:kurt} is reachable only through the third-order ($S\ge3$) reserve. The conditional-$\SLS$ gain, like the $\PMM_3$ gain, is an oracle-field quantity: a feasible version must estimate $\kappa_3(x)$ and $\mu_4(x)$ stratum-wise.

\section{Discussion and limitations}
\label{sec:discussion}

\paragraph{The contribution is structural}
The central claim is structural rather than numerical: optimal design for the higher-order
$\PMM$ estimators is a distinct problem, governed by $\Mg(\xi)$ and the transformed regressor
$\bh(x)=\gS(x)^{-1/2}\bef(x)$. A field constant on $\mathcal{X}$ makes $\Mg$ a positive scalar
multiple of $\bM$ and every standard criterion returns the $\OLS$ design; a non-constant field
can move the design, and whether it does depends on the criterion and the model
(Theorem~\ref{thm:dichotomy}). A second-order design reads a varying skewness field when the
estimator uses the common variance (the conditional $\SLS$ gains $+2.5\%$ on the \textsc{skew}
field) but no symmetric field; the symmetric, kurtosis-heteroscedastic regime is reachable only at
order $S\ge3$ (Theorem~\ref{thm:reserve}, Proposition~\ref{prop:csls}).

\paragraph{What is proved and what is illustrated}
The equivalence theorem, the degeneracy dichotomy, Theorem~\ref{thm:reserve} and
Proposition~\ref{prop:csls} are proved for a given reduction field. The link from $\Mg$ to an
estimator (Proposition~\ref{prop:adaptive}) is proved at every degree (Lemma~\ref{lem:gS}); outside the
symmetric class, with a free shape, the realised field is identically one. All of these results assume independent
errors, a supplied (known) reduction field and finite moments of order $2S$. They assume a constant variance; for the
stratify-and-pool estimator a varying variance changes only the weights (Remark~\ref{rem:variance}), while the
conditional $\SLS$ results need the common variance. Proposition~\ref{prop:plugin} gives the limit of a plug-in
design and Proposition~\ref{prop:asym} the cost of a small asymmetry; neither removes the known-field assumption from
the optimality theory. The numbers come from one worked
example: a quadratic dose--response model on $[0,1]$, two error families (symmetric Beta and centred
Gamma), a calibrated field treated as known, the $A$-criterion with a single $D$ comparison, and one
pilot fraction for the two-stage procedure. They show that the predicted separation occurs and how
large it is there; they do not show how large it is for other models, fields or criteria.

\paragraph{The gain is modest}
Under the $g_3$-weighted $A$-criterion the analytic efficiency of the $\PMM_3$-optimal design over
the $\OLS$-optimal design is $+4.7\%$ (ratio $1.047$), computed from the calibrated field. This
modest figure is the headline number. The optimal support shifts only mildly: the interior
dose moves from $0.50$ to $0.54$ and weight migrates off the top dose ($0.19\to0.13$). The
reason is that a fixed-degree polynomial estimator cannot use the hard edge of the uniform end
of the Beta$(a,a)$ family, so the kurtosis reserve $g_3(x)$ remains a regular,
$\sqrt{n}$-rate constant reduction down to $a=1$ rather than diverging. We do not
claim that cumulant-heteroscedastic design routinely buys double-digit precision. We
claim that it is a well-posed problem with a strictly positive answer where the second-order
theory has none.

\paragraph{Criterion-dependence ($A$ versus $D$)}
The dichotomy of Theorem~\ref{thm:dichotomy} does not depend on the criterion: in the
constant-$\gS$ case every criterion invariant under a positive scalar multiple of the
information matrix returns the $\OLS$ design, and otherwise the transported Kiefer--Wolfowitz
equivalence applies to any differentiable $\Psi$. The \emph{magnitude} of separation does depend
on the criterion. The headline gain above is the $A$-optimal one. Under $D$-optimality
(Table~\ref{tab:criterion}) the second-order separation is again $0$ and the third-order
separation positive, but the gain is only $+0.1\%$. The $D$-optimal design for the quadratic
holds the three support weights at $\tfrac13$, so the $g_3$ field can move it only by displacing
the interior point ($0.50\to0.52$); the recorded separation of $0.333$ buys almost nothing.
The $A$-criterion can also move weight off the top dose and onto the interior, which gives the
larger gain. The efficiency number must therefore be quoted per criterion.

\paragraph{Finite-sample behaviour and the $\gamma_6$ plug-in cost}
The reserve is an asymptotic property of $\gS$. The locally-adaptive $\PMM_3$ realises
it only when each high-reserve stratum carries enough samples to estimate the cumulants
that control the reduction. Degree-3 $\PMM$ couples to $\gamma_6$, and $\hat\gamma_6$ is noisy
in small strata, so the realised gain emerges with $N$ rather than at every sample size.
In the Monte Carlo $N$-sweep the point estimate of the total-MSE gain
is $0.989$ at $N=300$, $1.004$ at $N=600$, $1.054$ at $N=900$ and $1.075$ at $N=1200$. This is a pattern in point estimates, suggestive rather than confirmatory; we did not fit a joint trend. All four
$95\%$ confidence intervals still contain one (e.g.\ $[0.977,1.135]$ at $N=900$ and
$[0.998,1.150]$ at $N=1200$), so no single sample size yields a gain individually
distinguishable from unity in that sweep. The separate run of Table~\ref{tab:twostage} has narrower
intervals, all above one at $N=600$ to $2400$. The claim rests on that replication and on the analytic computation, and the sweep is exploratory: four sample sizes
locate a crossover but do not estimate a threshold. The pooled,
global-moment $\PMM_3$ has point estimates near one ($1.003,1.001,1.023,0.970$
across the sweep). No intervals were computed for them, so they are descriptive only and support no
contrast with the locally-adaptive estimator. In
practice the controlling high-order cumulants must be estimable stratum-wise: at degree $S$ this
requires finite moments of order $2S$ at every dose, and for heavier tails the field is undefined,
not merely noisy. That the first sweep does not recover the gain below $N=900$ is consistent with,
but does not isolate, the plug-in cost of $\hat\gamma_6$; the separate run recovers $1.035$ already
at $N=600$, so the location of the crossover is uncertain.

\paragraph{The reduction field is treated as known}
Both the optimal design and the inverse-variance weights of the locally-adaptive estimator use
the reduction field $\gS(x)$, so the reported gain is a \emph{locally-optimal} (oracle-field)
gain. It is the efficiency available once the field is known, the design-theoretic analogue of a
locally optimal design for a nonlinear model, whose optimal points depend on the unknown
parameters \citep{chernoff1953locally,atkinson2007optimum,dette2008optimaldose}. In a real dose--response study the
cumulant field is not known a priori. The natural feasible procedure has two stages, as in two-stage adaptive designs for nonlinear models \citep{dette2013twostage,lane2014information}: a pilot
allocation estimates the field stratum-wise, and the second stage allocates against the estimated
$\hat g_S(x)$. The price is the two-stage efficiency loss plus the $\hat\gamma_6$ plug-in cost of Remark~\ref{rem:threshold}.
For a fixed design the plug-in cost vanishes as $N\to\infty$ (Lemmas~\ref{lem:g3}(c) and~\ref{lem:gS}(b)). For the
design, Proposition~\ref{prop:plugin} gives the limit: a field estimate consistent on the whole grid and a pilot
fraction that shrinks make the plug-in design asymptotically efficient, while a fixed pilot fraction or a field
estimated only at the pilot doses leaves a loss that does not vanish. The proposition does not cover the realised
mean squared error on a data-chosen design, and we present the two-stage procedure as an exploratory
simulation, not as a validated implementation. Table~\ref{tab:twostage} shows that
the price is not small at the sample sizes we studied. With a one-third pilot, the design built at
$N=600$ to $2400$, evaluated under the true field, has an $A$-efficiency gain of $1.4\%$ to $2.0\%$, that is
$29\%$ to $43\%$ of the oracle gain. All pilot observations stay in the final design, and the procedure reads the field only at the three
pilot doses. Even with these values exact, its limiting gain is therefore $2.3\%$ ($48\%$ of the oracle gain);
pilot fractions of $0.2$, $0.1$ and $0.05$ raise it to $3.1\%$, $3.5\%$ and $3.5\%$, close to the
$3.5\%$ of a one-stage design built on the interpolated field. With the field known on the whole region the same pilot fractions would give $2.8\%$, $3.8\%$, $4.4\%$ and
$4.6\%$. A better pilot therefore spreads over more doses, not only over fewer observations. The
realised gain is a loss at $N=600$ and
fluctuates between $0.965$ and $1.080$ at $N=900$ to $2400$ without a monotone trend. The oracle-field gain should therefore be read as
the reserve the design can reach once the field is known, for instance from earlier studies of the same
assay, not as what a single study with an unknown field will obtain. A wrongly assumed field makes the
design and the pooling weights optimal for the wrong field, and the gain need not transfer; the
two-stage study is one instance of this, and a systematic study of misspecified level, slope and shape of
the field is not attempted here. Better pilot allocations, smoothing
of the estimated field and a fully sequential variant are left to future work. The present paper isolates the design-theoretic
object, the dichotomy and the third-order separation, for which the oracle-field gain
is the appropriate idealisation.

\paragraph{Symmetry must hold to within sampling error}
The stratify-and-pool $\PMM_3$ takes its reserve from the symmetry it imposes, and Proposition~\ref{prop:asym}
prices that restriction. A stratum gains over its mean only while its absolute skewness is below one asymptotic standard
error of its sample skewness, and a fixed skewness makes the root inconsistent, so that the gain turns into a loss
as $N$ grows (Corollary~\ref{cor:asym}). A pilot test of symmetry cannot certify the gain, because a skewness at
the break-even level is detected with low power. The design gain is a statement about symmetric strata. Where
the symmetry is in doubt, the free-shape stratum mean is the safe fit; its field is $\gS\equiv1$, and the design
reverts to the $\OLS$ one.

\paragraph{Choice of degree and an uncertain field}
Under symmetry an even degree adds nothing ($g_{2r}=g_{2r-1}$) and $g_{S+2}\le\gS$ (Lemma~\ref{lem:gS}(c)), but
each step up by two needs finite moments of two more orders and estimates of the cumulants up to that order in
every stratum. The degree is therefore an input of the design, to be chosen from the pilot: the smallest odd $S$
whose estimated field varies over the candidate support and whose cumulants the planned stratum sizes can
estimate. We did not develop a data-driven rule. Because $\Mg$ is an ordinary information matrix in $\bh$, designs
that are robust to an uncertain field, maximin-efficient over a set of fields or Bayesian over a prior on the
field, are classical problems in transformed regressors and can be computed by the same algorithms; we did not
compute them.

\paragraph{A finite-sample improvement through the basis}
Part of the $\gamma_6$ plug-in cost comes from the basis: it arises when the locally-adaptive
procedure estimates a basis-dependent reduction field, and in the two-stage procedure the same
estimate also chooses the design and the pooling weights, so the two costs are not cleanly
separable. The integer degree-3 polynomial used here is only the estimator inside the design
problem, and it can be replaced by the fractional, parametrically-adaptive transition-polynomial
(PATP) basis of \citet{zabolotnii2026patp}. Its continuous-order exponents trade off bias against
the variance of the high-order cumulant estimates and so may stabilise the per-stratum
reduction at smaller $N$; we have not tested this, and a lower sample-size requirement is a
hypothesis for future work. Such a change would alter the \emph{estimator inside} the design, not
the object of the paper: any basis, integer or fractional, enters only as the device that realises
$\gS(x)$, and the dichotomy and the equivalence theorem hold for whatever field it realises.

\section{Conclusion}
\label{sec:conclusion}

We have formulated optimal experimental design for the Polynomial Maximization Method when the
error shape varies with the dose at constant variance. The design results hold for a given reduction
field and every degree. Separately, the stratify-and-pool estimator has asymptotic covariance $\sigma^2\Mg(\xi)^{-1}$ at every degree under the
per-stratum conditions of Lemma~\ref{lem:gS}. In this locally-adaptive estimator a stratum with an
unrestricted shape realises no reduction, and a reduction below one arises only where the errors are
symmetric and the estimator imposes the symmetry. Two results carry the paper. The
degeneracy dichotomy (Theorem~\ref{thm:dichotomy}) shows that a non-constant reduction field is necessary for the design to move, but not sufficient by itself. The order dichotomy (Theorem~\ref{thm:reserve},
Proposition~\ref{prop:csls}) shows that a second-order design built on the common-variance
conditional $\SLS$ can follow a varying skewness, while the stratify-and-pool $\PMM_2$ cannot, and
that the symmetric, kurtosis-heteroscedastic regime is reachable only from order three. In the
worked example the third-order design reaches an oracle-field (known-field) $A$-efficiency of $1.047$ over the
$\OLS$-optimal design; the design built by a two-stage procedure with an estimated field reached only $1.014$ to $1.020$. A
plug-in design is asymptotically efficient only with a field estimate consistent on the whole region and a
vanishing pilot fraction (Proposition~\ref{prop:plugin}), and the third-order gain survives an asymmetry only while its absolute value
is below one standard error of the stratum's sample skewness (Proposition~\ref{prop:asym}). The first next step is
design under an unknown field: pilots spread over more doses, the error of the estimated field, robust designs over
a set of fields and a fully sequential allocation. Others are $D$- and $c$-optimal designs beyond the
single $D$ computation reported here, finite-sample behaviour with the fractional PATP basis of
\citet{zabolotnii2026patp}, and a proof that the Newton iteration of EstemPMM returns the root nearest
to $\bar y$ that Lemmas~\ref{lem:g3} and~\ref{lem:gS} analyse.

\appendix
\counterwithin*{equation}{section}
\counterwithin*{table}{section}
\renewcommand{\theequation}{\Alph{section}.\arabic{equation}}
\renewcommand{\thetable}{\Alph{section}.\arabic{table}}

\section{Proof of \texorpdfstring{Proposition~\ref{prop:adaptive}}{Proposition 1}}
\label{app:prop1}

Let $x_1,\dots,x_K$ be the support points of $\xi$, let $\mathbf{F}$ be the $K\times p$ matrix with
rows $\bef(x_k)^\top$, and write $\boldsymbol{\mu}=(\mu_1,\dots,\mu_K)^\top=\mathbf{F}\btheta$ and
$\hat{\boldsymbol{\mu}}=(\hat\mu_1,\dots,\hat\mu_K)^\top$. The stratum sizes $n_k=n_k(N)$ are integers with
$n_k/N\to w_k>0$; $n_k=Nw_k$ when these are integers. Write $g_k=\gS(x_k)\in(0,1]$ and
$\mathbf{W}=\mathrm{diag}(w_k/g_k)$. The pooling weights are
$\hat{\mathbf{W}}=\mathrm{diag}\{n_k/(N\hat g_k)\}$, where $\hat g_k=g_k$ for the known field of the
proposition; the argument below needs only $\hat g_k>0$ and $\hat g_k\to g_k$ in probability, so it
covers weights computed from a strictly positive consistent estimate of the field as well, including
one computed from the same data, since nothing else about $\hat g_k$ is used (in the simulations the
estimates are clipped to $[0.05,1]$, and since the true field is at least $0.30$ on the \textsc{kurt} region
the lower clip does not bind in the limit). The number $K$ of strata stays fixed as $N\to\infty$. The argument concerns a fixed design; it says nothing about designs whose support or allocation is itself chosen from the estimated field, as in the two-stage procedure of Section~\ref{subsec:kurt}.

\emph{Step 1 (an exact identity).} Since $\boldsymbol{\mu}=\mathbf{F}\btheta$ holds exactly, the generalised
least-squares estimate $\hat\btheta=(\mathbf{F}^\top\hat{\mathbf{W}}\mathbf{F})^{-1}\mathbf{F}^\top\hat{\mathbf{W}}\hat{\boldsymbol{\mu}}$ satisfies
\begin{equation}
\label{eq:app-identity}
\hat\btheta-\btheta=\hat{\mathbf{A}}\,(\hat{\boldsymbol{\mu}}-\boldsymbol{\mu}),\qquad
\hat{\mathbf{A}}=(\mathbf{F}^\top\hat{\mathbf{W}}\mathbf{F})^{-1}\mathbf{F}^\top\hat{\mathbf{W}},
\end{equation}
whenever $\mathbf{F}^\top\hat{\mathbf{W}}\mathbf{F}$ is invertible. It always is: $\hat{\mathbf{W}}$ is diagonal with
positive entries and $\mathbf{F}$ has full column rank $p$ because $\bM(\xi)=\mathbf{F}^\top\mathrm{diag}(w_k)\mathbf{F}$ is
non-singular, so $\mathbf{F}^\top\hat{\mathbf{W}}\mathbf{F}$ is positive definite.

\emph{Step 2 (the limit of the weights).} $\mathbf{F}^\top\mathbf{W}\mathbf{F}=\sum_k (w_k/g_k)\,\bef(x_k)\bef(x_k)^\top=\Mg(\xi)$
by \eqref{eq:Mg}. Because $g_k\le1$, $\Mg(\xi)-\bM(\xi)=\sum_k w_k(g_k^{-1}-1)\bef(x_k)\bef(x_k)^\top$ is
positive semidefinite, so $\Mg(\xi)$ is positive definite whenever $\bM(\xi)$ is. Since $\hat{\mathbf{W}}\to\mathbf{W}$
in probability, $\hat{\mathbf{A}}\to\mathbf{A}=\Mg(\xi)^{-1}\mathbf{F}^\top\mathbf{W}$ in probability, by continuity of matrix inversion.

\emph{Step 3 (joint convergence of the stratum estimates).} Put $\mathbf{Z}_N=\sqrt N(\hat{\boldsymbol{\mu}}-\boldsymbol{\mu})$. Its
$k$-th coordinate is $(N/n_k)^{1/2}\sqrt{n_k}(\hat\mu_k-\mu_k)$, which converges in distribution to
$N(0,\sigma^2g_k/w_k)$ by Assumption~\ref{ass:stratum} and Slutsky's theorem. The strata are independent, so the
characteristic function of $\mathbf{Z}_N$ is the product of the coordinate characteristic functions, and
$\mathbf{Z}_N\xrightarrow{d}N(\mathbf{0},\sigma^2\mathbf{W}^{-1})$.

\emph{Step 4 (conclusion).} By \eqref{eq:app-identity}, $\sqrt N(\hat\btheta-\btheta)=\hat{\mathbf{A}}\mathbf{Z}_N$, and
Slutsky's theorem gives the limit $N(\mathbf{0},\sigma^2\mathbf{A}\mathbf{W}^{-1}\mathbf{A}^\top)$, with
\[
\mathbf{A}\mathbf{W}^{-1}\mathbf{A}^\top
=\Mg(\xi)^{-1}\mathbf{F}^\top\mathbf{W}\mathbf{W}^{-1}\mathbf{W}\mathbf{F}\,\Mg(\xi)^{-1}
=\Mg(\xi)^{-1}. \qquad\square
\]

Two consequences are used in the paper. First, among all pooled estimators $\mathbf{L}\hat{\boldsymbol{\mu}}$ with
a fixed matrix $\mathbf{L}$, or one converging in probability to a deterministic matrix, such that $\mathbf{L}\mathbf{F}=\mathbf{I}_p$, the limiting covariance $\sigma^2\mathbf{L}\mathbf{W}^{-1}\mathbf{L}^\top$ exceeds
$\sigma^2\Mg(\xi)^{-1}$ in the Loewner order (Aitken's theorem applied to the limiting covariance
$\sigma^2\mathbf{W}^{-1}$), so the weights $n_k/\gS(x_k)$ are asymptotically best within that class of
fixed-design linear stratify-and-pool estimators. The statement concerns pooling only; it says nothing about the conditional-$\SLS$ information or about adaptive designs. The statement does not cover nonlinear or other data-dependent combinations. Second, when $\xi$ has exactly $K=p$ support points,
$\mathbf{F}$ is square and invertible, $\hat\btheta=\mathbf{F}^{-1}\hat{\boldsymbol{\mu}}$ for every choice of weights, and the
field enters the estimator only through the per-stratum estimates; the optimal designs reported in
Section~\ref{sec:numerical} have three support points.

\emph{Numerical check (not part of the proof).} The algebra of Steps~2 and 4 is checked symbolically, and
the limit by a Monte Carlo with per-stratum $\PMM_3$ on Beta errors at four strata ($N=8000$, $4000$
replications): the diagonal of $N\,\widehat{\Cov}(\hat\btheta)$ lies within $3\%$ of that of
$\sigma^2\Mg(\xi)^{-1}$. For scale, a variance estimated from $4000$
Gaussian replicates has a relative standard error of about $\sqrt{2/4000}\approx2.2\%$. The check is one of
consistency, not a test at a $3\%$ threshold, and the $3\%$ includes the finite-$N$ bias.

\section{Proof of \texorpdfstring{Proposition~\ref{prop:asym}}{the local-asymmetry proposition}}
\label{app:asym}

Write $\psi_\kappa(t)=t(\kappa-t^2)$ as in Lemma~\ref{lem:g3}, $\kappa^\ast$ for its optimal coefficient at $P$, and
$t=\hat\mu-\mu$. Since $\sup_n\E|e_{n,1}|^{6+\delta}<\infty$, the powers $|e_{n,1}|^r$, $r\le6$, are uniformly integrable, and
the weak law of large numbers for triangular arrays gives $n^{-1}\sum_ie_{n,i}^r-\E e_{n,1}^r\to0$ in probability. The
sample central moments therefore converge to the central moments of $P$, and $\hat\kappa\to\kappa^\ast$, with
$\kappa^\ast-3m_2=m_2D/\gamma_4\neq0$ (proof of Lemma~\ref{lem:g3}(b)).

\emph{Existence and consistency of the root.} With $m_{r,n}$ the central moments of $P_n$,
$\E\psi_\kappa(e_{n,1}-s)=-m_{3,n}-s(\kappa-3m_{2,n})+s^3$. As $m_{3,n}\to0$ and $m_{2,n}\to m_2$, for every small
$\delta'>0$ this function takes opposite signs at $s=\pm\delta'$ for all large $n$, uniformly for $\kappa$ near
$\kappa^\ast$. The argument of Lemma~\ref{lem:g3}(a) and (c), with the law of large numbers for triangular arrays,
then shows that the root nearest to $\bar y$ is consistent.

\emph{Expansion.} As in Lemma~\ref{lem:g3}(a), exactly,
\begin{align*}
0=\frac1{\sqrt n}\sum_i\psi_{\hat\kappa}(e_{n,i})&-\sqrt n\,t\,\frac1n\sum_i(\hat\kappa-3e_{n,i}^2)\\
&-3\sqrt n\,t^2\,\bar e_n+\sqrt n\,t^3 .
\end{align*}
By linearity in $\kappa$ the first term equals
\[
n^{-1/2}\textstyle\sum_i\psi_{\kappa^\ast}(e_{n,i})+(\hat\kappa-\kappa^\ast)\,n^{-1/2}\sum_ie_{n,i},
\]
and the second part is $o_p(1)$ because $\E e_{n,1}=0$ and the variances are bounded. Since
$\E\psi_{\kappa^\ast}(e_{n,1})=-m_{3,n}$ and $\sqrt n\,m_{3,n}=\sqrt n\,\gamma_3(P_n)\,m_{2,n}^{3/2}\to h\,m_2^{3/2}$,
\begin{align*}
\frac1{\sqrt n}\sum_i\psi_{\kappa^\ast}(e_{n,i})
&=\frac1{\sqrt n}\sum_i\big\{\psi_{\kappa^\ast}(e_{n,i})-\E\psi_{\kappa^\ast}(e_{n,1})\big\}-\sqrt n\,m_{3,n}\\
&\xrightarrow{\,d\,}N\big(-h\,m_2^{3/2},\ \E_P\psi_{\kappa^\ast}^2\big)
\end{align*}
by the Lyapunov central limit theorem: $|\psi_{\kappa^\ast}(e)|^{2+\delta/3}\le C(1+|e|^{6+\delta})$ and
$\E\psi_{\kappa^\ast}(e_{n,1})^2\to\E_P\psi_{\kappa^\ast}^2$ by the convergence of the moments. The average in the second term
tends to $\kappa^\ast-3m_2\neq0$, and consistency gives $\sqrt n\,t=O_p(1)$ and the last two terms $o_p(1)$, as in
Lemma~\ref{lem:g3}(a). Hence
\begin{align*}
\sqrt n\,t&\xrightarrow{\,d\,}N\Big(\frac{-h\,m_2^{3/2}}{\kappa^\ast-3m_2},\ \frac{\E_P\psi_{\kappa^\ast}^2}{(\kappa^\ast-3m_2)^2}\Big)\\
&=N\Big(-\sqrt{m_2}\,\frac{\gamma_4}{D}\,h,\ m_2\,g_3\Big),
\end{align*}
using $\kappa^\ast-3m_2=m_2D/\gamma_4$ and Lemma~\ref{lem:g3}(b) for the variance.

\emph{Mean squared error and the skewness.} The second moment of the limit is
$m_2\{g_3+\gamma_4^2h^2/D^2\}$, and $1-g_3=\gamma_4^2/D$ by \eqref{eq:g3}, which gives \eqref{eq:asym-mse}. Under $P$ the
sample skewness $\hat m_3/\hat m_2^{3/2}$ has, since $m_3=0$, the influence function $\varepsilon^3-3\varepsilon$ with
$\varepsilon=e/\sqrt{m_2}$, whose variance is $\E\varepsilon^6-6\E\varepsilon^4+9=D$. \qed

\section{The third-order location reduction}
\label{app:g3}

This appendix derives $g_3$ for a symmetric error law and shows that co-estimating the scale does not
change it, which is the fact behind Theorem~\ref{thm:reserve}(ii) at $S=3$. Let $y_i=\mu+e_i$,
$i=1,\dots,n$, with $e_i$ i.i.d., symmetric about zero, $\E e^6<\infty$, central moments $m_r=\E e^r$, and
standardised cumulants $\gamma_4=m_4/m_2^2-3$ and $\gamma_6=m_6/m_2^3-15\,m_4/m_2^2+30$. Write
$\varepsilon=e/\sqrt{m_2}$ and $\mathrm{He}_3(t)=t^3-3t$.

\begin{lemma}[Degree-3 location reduction]
\label{lem:g3}
Assume $\gamma_4\neq0$ and that the law of $e$ is not concentrated on three or fewer points. For
$\kappa\in\R$ let $\psi_\kappa(t)=t(\kappa-t^2)$, the degree-3 combination of the first and third powers.
\begin{enumerate}
\item[(a)] For fixed $\kappa\ne3m_2$, with probability tending to one the equation $\sum_i\psi_\kappa(y_i-\mu)=0$
has a root within any fixed distance of the true $\mu$; let $\hat\mu_\kappa$ be the root nearest to $\bar y$.
Then $\hat\mu_\kappa$ is $\sqrt n$-consistent and $\sqrt n(\hat\mu_\kappa-\mu)\xrightarrow{d}N\big(0,\ m_2\,r(\kappa)\big)$ with
$r(\kappa)=(\kappa^2m_2-2\kappa m_4+m_6)/\{m_2(3m_2-\kappa)^2\}$.
\item[(b)] $r$ is minimised at $\kappa^\ast=(m_6-3m_2m_4)/(m_4-3m_2^2)$, and
\begin{equation}
\label{eq:g3}
g_3:=r(\kappa^\ast)=1-\frac{\gamma_4^2}{6+9\gamma_4+\gamma_6}
=1-\operatorname{Corr}^2\!\big(\varepsilon,\mathrm{He}_3(\varepsilon)\big)\in(0,1).
\end{equation}
As $\kappa\to\infty$, $r(\kappa)\to1$, the sample mean.
\item[(c)] Let $\hat\kappa$ be $\kappa^\ast$ evaluated at the sample central moments about $\bar y$ when its
denominator $\hat m_4-3\hat m_2^2$ is non-zero, and let $\hat\mu$ be the root of
$\sum_i\psi_{\hat\kappa}(y_i-\mu)=0$ nearest to $\bar y$, with $\hat m_r=n^{-1}\sum_i(y_i-\bar y)^r$; on the event $\hat m_4=3\hat m_2^2$ put
$\hat\mu=\bar y$. Then
$\sqrt n(\hat\mu-\mu)\xrightarrow{d}N(0,m_2\,g_3)$: estimating the scale and the shape costs nothing at first
order.
\end{enumerate}
\end{lemma}

\begin{proof}
(a) \emph{Existence and consistency of the root.} By symmetry, $\E\psi_\kappa(e-t)=-t(\kappa-3m_2)+t^3$, which for
every small $\delta>0$ takes opposite signs at $t=\pm\delta$ because $\kappa\ne3m_2$. By the law of large numbers the
sample average $n^{-1}\sum_i\psi_\kappa(y_i-\mu-t)$ converges at $t=\pm\delta$ to these values, so with probability
tending to one it changes sign on $[-\delta,\delta]$ and, being continuous in $t$, has a root $\mu+t_n$ with
$|t_n|\le\delta$. The root nearest to $\bar y$ is no farther from $\bar y$ than this one, so it lies within
$2|\bar y-\mu|+\delta$ of $\mu$. Since $\delta$ is arbitrary and $\bar y\to\mu$, $\hat\mu_\kappa$ is consistent.
\emph{Expansion.} Because $\psi_\kappa$ is a cubic polynomial, the expansion about the true value is exact: with
$t=\hat\mu_\kappa-\mu$,
\begin{align*}
0=\frac1{\sqrt n}\sum_i\psi_\kappa(e_i-t)
&=\frac1{\sqrt n}\sum_i\psi_\kappa(e_i)-\sqrt n\,t\,\frac1n\sum_i(\kappa-3e_i^2)\\
&\quad-3\sqrt n\,t^2\,\bar e+\sqrt n\,t^3 .
\end{align*}
By symmetry $\E\psi_\kappa(e)=\kappa m_1-m_3=0$, and $\E\psi_\kappa(e)^2=\kappa^2m_2-2\kappa m_4+m_6<\infty$, so the
first term is asymptotically $N(0,\E\psi_\kappa^2)$. The average in the second term tends to $\kappa-3m_2\ne0$.
Collecting the terms in $\sqrt n\,t$ gives $\sqrt n\,t\,\{\kappa-3m_2+o_p(1)+O_p(t)\}=O_p(1)$, and $t=o_p(1)$ by
consistency, so $\sqrt n\,t=O_p(1)$; the last two terms are then $o_p(1)$. Solving for $\sqrt n\,t$ gives the limit with
variance $\E\psi_\kappa^2/(3m_2-\kappa)^2=m_2\,r(\kappa)$.

(b) Setting $dr/d\kappa=0$ gives the single root $\kappa^\ast$, which is EstemPMM's $\kappa$. Substituting
$m_4=(3+\gamma_4)m_2^2$ and $m_6=(\gamma_6+15\gamma_4+15)m_2^3$ gives the middle expression of \eqref{eq:g3}.
For the last, $\E\varepsilon^2=1$, $\E[\varepsilon\,\mathrm{He}_3(\varepsilon)]=\gamma_4$ and
$\E\mathrm{He}_3(\varepsilon)^2=6+9\gamma_4+\gamma_6$, and both $\varepsilon$ and $\mathrm{He}_3(\varepsilon)$ have mean zero.
The correlation is not $\pm1$ because $\varepsilon^3$ is an affine function of $\varepsilon$ only for a law on at most
three points ($\{0,\pm c\}$), which is excluded; it is not zero because $\gamma_4\ne0$. Hence $0<g_3<1$. At
$\kappa^\ast$ the second derivative of $r$ equals $2\gamma_4^4/(6+9\gamma_4+\gamma_6)^3>0$ for $m_2=1$, so
the stationary point is a local minimum. It is the global one. The numerator of $r$ is
$\E[e^2(\kappa-e^2)^2]>0$, because a law on which $e^2(\kappa-e^2)^2$ vanishes almost surely sits on at
most three points; hence $r\to+\infty$ at the excluded pole $\kappa=3m_2$, and $r(\kappa)\to1$ as
$\kappa\to\pm\infty$ by direct computation. On the side of the pole that does not contain $\kappa^\ast$, $r$
has no stationary point and so decreases monotonically from $+\infty$ towards $1$, staying above one; on
the other side its only stationary point is $\kappa^\ast$, with $r(\kappa^\ast)=g_3<1$. Finally
$\kappa^\ast-3m_2=m_2\,(6+9\gamma_4+\gamma_6)/\gamma_4\ne0$, so (a) applies at $\kappa^\ast$.

(c) The function $\psi_\kappa$ is affine in $\kappa$ with $\partial\psi_\kappa(t)/\partial\kappa=t$. Hence, exactly,
$n^{-1/2}\sum_i\psi_{\hat\kappa}(e_i)=n^{-1/2}\sum_i\psi_{\kappa^\ast}(e_i)+(\hat\kappa-\kappa^\ast)\,n^{-1/2}\sum_ie_i$.
The sample central moments are consistent under $\E e^6<\infty$ and the population denominator
$m_4-3m_2^2=\gamma_4m_2^2$ is non-zero, so the event $\hat m_4=3\hat m_2^2$ has probability tending to zero and
does not affect the limit, and $\hat\kappa-\kappa^\ast=o_p(1)$, while
$n^{-1/2}\sum_ie_i=O_p(1)$; the extra term is $o_p(1)$. The same holds for the average
$n^{-1}\sum_i(\hat\kappa-3e_i^2)\to\kappa^\ast-3m_2$, and both the existence argument and the expansion of (a) go
through with $\hat\kappa$ in place of $\kappa$; in the former, $n^{-1}\sum_i\psi_{\hat\kappa}(e_i\mp\delta)$ differs from
$n^{-1}\sum_i\psi_{\kappa^\ast}(e_i\mp\delta)$ by $(\hat\kappa-\kappa^\ast)\,n^{-1}\sum_i(e_i\mp\delta)=o_p(1)$. The reason is $\E[\partial\psi_\kappa(e)/\partial\kappa]=\E e=0$: the direction in which the scale and the
shape enter is orthogonal to the location score, which is the decoupling of the squared direction under
symmetry that Theorem~\ref{thm:reserve}(ii) cites \citep[Theorem~3]{zabolotnii2026slspmm}.
\end{proof}

\emph{Implementation note.} The lemma concerns the root nearest to $\bar y$; a cubic equation can have up to
three roots. EstemPMM's intercept-only degree-3 fit computes $\hat m_2,\hat m_4,\hat m_6$ from the least-squares
residuals, here $y_i-\bar y$, forms $\hat\kappa$ as in (c), and runs a Newton iteration started at $\bar y$ (we
checked this in version~0.5.0). We did not prove that the iteration always returns the nearest root; the
calibration treats its output as that root.

For $S=2$ the analogous statement fails in the opposite direction: the degree-2 system of a single
stratum is exactly identified and returns the sample mean (Lemma~\ref{lem:loc2}). For $S=3$ and a
symmetric law not concentrated on three or fewer points, with the symmetry imposed, the location keeps its asymptotic reduction $g_3<1$ whenever $\gamma_4\ne0$.

\paragraph{The Beta family} For a centred $\mathrm{Beta}(a,a)$ variate the central moments are
$\E(X-\tfrac12)^{2j}=4^{-j}\prod_{i=0}^{j-1}(\tfrac12+i)/(a+\tfrac12+i)$, which give, with $u=2a+3$,
\[
\gamma_4=-\frac6u,\qquad \gamma_6=\frac{240}{u(u+2)},\qquad
g_3(a)=1-\frac{6(u+2)}{u\,(u^2-7u+22)} .
\]
At $a=1$ (uniform) $\gamma_4=-6/5$, $\gamma_6=48/7$ and $g_3=0.300$. Differentiating in $u$ shows that $g_3(a)$
is strictly increasing in $a>0$, so under the \textsc{kurt} field $a(x)=8-7x$ the asymptotic $g_3(x)$ is strictly
decreasing in the dose and non-constant on every design with two or more support points. Table~\ref{tab:g3closed} sets the
closed form beside the calibrated values of Table~\ref{tab:calib_kurt}. The calibrated values, from samples of size
$400$ with co-estimated scale and shape, exceed the asymptotic ones by up to $0.027$ (at $a=8$), with one
value below it ($a=1.6$, by $0.011$); this is consistent with the finite-sample plug-in cost of Remark~\ref{rem:threshold}
plus Monte Carlo error. We did not decompose the gap; root selection and the Newton iteration may also
contribute. Replacing the calibrated field by the closed-form field leaves the $A$-optimal
$\PMM_3$ design unchanged at the grid resolution ($\{0\!:\!0.34,\ 0.54\!:\!0.53,\ 1\!:\!0.13\}$) and moves the
analytic $A$-efficiency gain from $1.047$ to $1.048$. Every identity in this appendix is verified symbolically
by a script in the code supplement (see Code availability).

\begin{table}[t]
\centering
\caption{The closed-form and calibrated third-order reductions agree to within $0.03$ and both fall as $a$
decreases toward the uniform law. Standardised fourth and sixth cumulants $\gamma_4,\gamma_6$ of
$\mathrm{Beta}(a,a)$, the asymptotic $g_3$ of \eqref{eq:g3}, and the calibration estimate $\hat g_3$ of
Table~\ref{tab:calib_kurt}: the Monte Carlo variance of the EstemPMM degree-3 location estimate divided by that of the sample mean, at sample size $n=400$ ($M_{\mathrm{cal}}=2500$ replications). The calibration estimate is a noisy finite-sample variance ratio and can exceed one, as at $a=8$; the asymptotic $g_3$ cannot.}
\label{tab:g3closed}
\begin{tabular}{rrrrr}
\toprule
Beta shape $a$ & $\gamma_4$ & $\gamma_6$ & $g_3$ closed form & $\hat g_3$ calibrated ($n=400$) \\
\midrule
8.0 & $-0.316$ & 0.602 & 0.973 & 1.001 \\
5.0 & $-0.462$ & 1.231 & 0.931 & 0.946 \\
3.5 & $-0.600$ & 2.000 & 0.862 & 0.881 \\
2.5 & $-0.750$ & 3.000 & 0.750 & 0.768 \\
2.0 & $-0.857$ & 3.810 & 0.649 & 0.659 \\
1.6 & $-0.968$ & 4.721 & 0.534 & 0.524 \\
1.3 & $-1.071$ & 5.639 & 0.425 & 0.446 \\
1.0 & $-1.200$ & 6.857 & 0.300 & 0.313 \\
\bottomrule
\end{tabular}

\end{table}

\section{The per-stratum location reduction at every degree}
\label{app:gS}

This appendix proves Assumption~\ref{ass:stratum} at every degree $S$. The coefficient \eqref{eq:gS-odd} defined below is the odd-power coefficient; part (b) of the lemma identifies it as the reduction for a symmetric law with the symmetry imposed, while part (a) gives the unit reduction of an unrestricted shape. As in \appref{app:g3}, let
$y_i=\mu+e_i$, $i=1,\dots,n$, with $e_i$ i.i.d., $\E e=0$, $\E e^{2S}<\infty$, central moments $m_r=\E e^r$, and a
law not concentrated on $S$ or fewer points; $\hat m_r=n^{-1}\sum_i(y_i-\bar y)^r$. Write $\mathcal{O}_S$ for the
odd integers in $\{1,\dots,S\}$, $S'=\max\mathcal{O}_S$, and
\begin{equation}
\label{eq:gS-odd}
\mathbf{F}=(m_{i+j})_{i,j\in\mathcal{O}_S},\qquad \mathbf{b}=(j\,m_{j-1})_{j\in\mathcal{O}_S},\qquad
\gS:=\frac{1}{m_2\,\mathbf{b}^\top\mathbf{F}^{-1}\mathbf{b}} .
\end{equation}

\begin{lemma}[Degree-$S$ location reduction]
\label{lem:gS}
Let $y_i=\mu+e_i$ with $e_1,\dots,e_n$ i.i.d., $\E e=0$, $\E e^{2S}<\infty$, and a law of $e$ that is not concentrated
on $S$ or fewer points.
\begin{enumerate}
\item[(a)] \emph{Unrestricted shape.} For every $S\ge1$ and every coefficient vector $\mathbf{k}\in\R^S$, fixed or
data-dependent, the degree-$S$ equation with the shape estimated in the stratum,
\[
\sum_{i=1}^n\sum_{j=1}^S k_j\big\{(y_i-\mu)^j-\hat m_j\big\}=0,\qquad \hat m_1=0,
\]
has the root $\mu=\bar y$. The same holds for the exactly identified moment system
$\E[(y-\mu)^j-m_j]=0$, $j=1,\dots,S$, in the unknowns $(\mu,m_2,\dots,m_S)$, whose sample version has the unique
solution $\bar y$ and the $\hat m_j$. The root nearest to $\bar y$ is $\bar y$ itself, and since $\E e^2<\infty$ the central limit theorem gives
$\sqrt n(\bar y-\mu)\xrightarrow{d}N(0,m_2)$; so Assumption~\ref{ass:stratum} holds with $\gS=1$ for every such law and
every $S$.
\item[(b)] \emph{Symmetric law, symmetry imposed.} Let the law of $e$ be symmetric about zero. Then $\mathbf{F}$ is
positive definite. For $\mathbf{k}\in\R^{|\mathcal{O}_S|}$ put $\psi_{\mathbf{k}}(t)=\sum_{j\in\mathcal{O}_S}k_jt^j$, let
$\hat{\mathbf{k}}=\hat{\mathbf{F}}^{-1}\hat{\mathbf{b}}$ be $\mathbf{k}^\ast=\mathbf{F}^{-1}\mathbf{b}$ evaluated at the
$\hat m_r$ when $\hat{\mathbf{F}}$ is non-singular, and let $\hat\mu$ be the root of
$\sum_i\psi_{\hat{\mathbf{k}}}(y_i-\mu)=0$ nearest to $\bar y$ (and $\hat\mu=\bar y$ when $\hat{\mathbf{F}}$ is singular).
Then $\sqrt n(\hat\mu-\mu)\xrightarrow{d}N(0,\,m_2\,\gS)$ with $\gS$ of \eqref{eq:gS-odd}, and
$\mathbf{k}^\ast$ minimises the asymptotic variance $\mathbf{k}^\top\mathbf{F}\mathbf{k}/(\mathbf{b}^\top\mathbf{k})^2$
over all odd polynomials of degree at most $S$ with non-zero sensitivity $\E\psi'_{\mathbf{k}}(e)=\mathbf{b}^\top\mathbf{k}\neq0$.
\item[(c)] \emph{Properties of the coefficient.} For the coefficient $\gS$ of \eqref{eq:gS-odd}, which is the
reduction of (b) and not the unit reduction of (a): $0<\gS\le1$; $g_{2r}=g_{2r-1}$; $\gS=1$ if and only if
$m_{j+1}=j\,m_2\,m_{j-1}$ for every odd $j$ with $3\le j\le S$; and $g_3$ equals the right side of \eqref{eq:g3},
which is $\min_\kappa r(\kappa)$ of Lemma~\ref{lem:g3} when $\gamma_4\neq0$ and equals one, attained by
$\psi(t)=t$, when $\gamma_4=0$. If the hypotheses also hold at degree $S+2$ ($\E e^{2S+4}<\infty$ and a law not
concentrated on $S+2$ or fewer points), then $g_{S+2}\le\gS$.
\end{enumerate}
\end{lemma}

\begin{proof}
(a) At $\mu=\bar y$ each bracket averages to $n^{-1}\sum_i(y_i-\bar y)^j-\hat m_j=0$ by the definition of $\hat m_j$,
for $j\ge2$, and to $n^{-1}\sum_i(y_i-\bar y)=0$ for $j=1$; the left side therefore vanishes whatever $\mathbf{k}$ is. The
moment system has as many equations as unknowns, and its sample version is solved by $\bar y$ and the $\hat m_j$, so
every non-singular combination of its equations has that root, as in the proof of Lemma~\ref{lem:loc2}.

(b) \emph{The matrix.} If $\mathbf{c}^\top\mathbf{F}\mathbf{c}=0$ then $\E p(e)^2=0$ for the odd polynomial
$p(t)=\sum_{j\in\mathcal{O}_S}c_jt^j=t\,q(t^2)$ with $\deg q\le(S'-1)/2$. If $\mathbf{c}\neq\mathbf{0}$, $p$ has at most
$S'\le S$ real zeros, and $e$ would be concentrated on them; this is excluded, so $\mathbf{F}\succ0$. Since $b_1=1$,
$\mathbf{b}\neq\mathbf{0}$ and $\mathbf{b}^\top\mathbf{k}^\ast=\mathbf{b}^\top\mathbf{F}^{-1}\mathbf{b}>0$.

\emph{The equation is the degree-$S$ PMM equation.} Kunchenko's degree-$S$ coefficients, those that minimise
the asymptotic variance of the root of a degree-$S$ moment equation for the location \citep{kunchenko2002polynomial},
solve
\[
\sum_{j=1}^S k_j\,\mathrm{Cov}(e^i,e^j)=i\,m_{i-1},\qquad i=1,\dots,S,
\]
where the right side is the derivative with respect to $\mu$ of the $i$-th moment of $y=\mu+e$ about a fixed
origin, evaluated at the origin $\mu$. Under symmetry every
odd moment vanishes, so $\mathrm{Cov}(e^i,e^j)=0$ when $i+j$ is odd, and the right side vanishes for even $i$. The
system splits: the even-index block has a zero right side, so zero coefficients on the even powers solve it, and
the odd-index block is $\mathbf{F}\mathbf{k}=\mathbf{b}$, since $\mathrm{Cov}(e^i,e^j)=m_{i+j}$ for odd $i,j$. With the even
powers absent their centring constants drop out, and the odd constants are the known zeros; this is what
imposing the symmetry means.

\emph{Existence and consistency of a root.} Because $\psi_{\mathbf{k}}$ is a polynomial,
\[
\E\psi_{\mathbf{k}}(e-t)=\sum_{r=0}^{S'}\frac{(-t)^r}{r!}\,\E\psi^{(r)}_{\mathbf{k}}(e)
\]
is a polynomial in $t$. Its constant term is $\E\psi_{\mathbf{k}}(e)=0$ by symmetry, and its linear term is
$-t\,\E\psi'_{\mathbf{k}}(e)=-t\,\mathbf{b}^\top\mathbf{k}$.
At $\mathbf{k}^\ast$ the slope is negative, so for every small $\delta>0$ the function takes opposite signs at
$t=\pm\delta$. The argument of Lemma~\ref{lem:g3}(a) then applies verbatim: by the law of large numbers the sample
average changes sign on $[-\delta,\delta]$ with probability tending to one, and the root nearest to $\bar y$ is
consistent.

\emph{Expansion.} With $t=\hat\mu-\mu$ the expansion is exact,
\[
0=\frac1{\sqrt n}\sum_i\psi_{\mathbf{k}}(e_i)-\sqrt n\,t\,\frac1n\sum_i\psi'_{\mathbf{k}}(e_i)
+\sum_{r=2}^{S'}\frac{(-1)^r\sqrt n\,t^r}{r!}\,\frac1n\sum_i\psi^{(r)}_{\mathbf{k}}(e_i).
\]
The first term is asymptotically $N(0,\E\psi_{\mathbf{k}}(e)^2)$ with $\E\psi_{\mathbf{k}}(e)^2=\mathbf{k}^\top\mathbf{F}\mathbf{k}<\infty$,
since $\E e^{2S'}<\infty$. The averages in the remaining terms converge to finite limits, the one in the second to
$\mathbf{b}^\top\mathbf{k}\neq0$. As in Lemma~\ref{lem:g3}(a), consistency gives $\sqrt n\,t=O_p(1)$, the terms with
$r\ge2$ are $o_p(1)$, and $\sqrt n\,t\to N\big(0,\mathbf{k}^\top\mathbf{F}\mathbf{k}/(\mathbf{b}^\top\mathbf{k})^2\big)$. By the
Cauchy--Schwarz inequality in the inner product of $\mathbf{F}$,
$(\mathbf{b}^\top\mathbf{k})^2\le(\mathbf{k}^\top\mathbf{F}\mathbf{k})(\mathbf{b}^\top\mathbf{F}^{-1}\mathbf{b})$, with equality at
$\mathbf{k}^\ast$; the minimal variance is $1/(\mathbf{b}^\top\mathbf{F}^{-1}\mathbf{b})=m_2\,\gS$.

\emph{Plug-in.} The function $\psi_{\mathbf{k}}$ is linear in $\mathbf{k}$, so exactly
$n^{-1/2}\sum_i\psi_{\hat{\mathbf{k}}}(e_i)=n^{-1/2}\sum_i\psi_{\mathbf{k}^\ast}(e_i)
+\sum_{j\in\mathcal{O}_S}(\hat k_j-k^\ast_j)\,n^{-1/2}\sum_ie_i^j$. For odd $j$, $\E e^j=0$ and $\Var e^j=m_{2j}<\infty$,
so $n^{-1/2}\sum_ie_i^j=O_p(1)$, while $\hat{\mathbf{k}}-\mathbf{k}^\ast=o_p(1)$ because the sample central moments of
order up to $2S'$ are consistent and $\mathbf{F}\succ0$; the extra term is $o_p(1)$. The event that $\hat{\mathbf{F}}$ is
singular has probability tending to zero. The same substitution leaves the derivative average and the existence
argument unchanged at first order, exactly as in Lemma~\ref{lem:g3}(c). In the existence argument,
$n^{-1}\sum_i\psi_{\hat{\mathbf{k}}}(e_i\mp\delta)$ differs from $n^{-1}\sum_i\psi_{\mathbf{k}^\ast}(e_i\mp\delta)$ by
$\sum_{j\in\mathcal{O}_S}(\hat k_j-k^\ast_j)\,n^{-1}\sum_i(e_i\mp\delta)^j=o_p(1)$, since each average converges to a finite
limit; so the sign change on $[-\delta,\delta]$, and with it the consistency of the root nearest to $\bar y$, survives
the plug-in. The reason is again
$\E[\partial\psi_{\mathbf{k}}(e)/\partial k_j]=\E e^j=0$ for odd $j$: the coefficients, and with them the scale and the
shape, enter in directions orthogonal to the location score.

(c) Positivity follows from $\mathbf{F}\succ0$. Restricting $\mathbf{k}$ to the first coordinate gives the sample mean,
with variance $m_2$, so the minimum in (b) is at most $m_2$ and $\gS\le1$. Equality holds if and only if
$\mathbf{k}^\ast$ is proportional to the first unit vector, that is, if and only if $\mathbf{F}\mathbf{e}_1\propto\mathbf{b}$; the
first coordinate fixes the constant at $m_2$, and the others read $m_{j+1}=j\,m_2\,m_{j-1}$. For even $S$ the set
$\mathcal{O}_S$ equals $\mathcal{O}_{S-1}$. When the hypotheses hold at degree $S+2$, the degree-$(S+2)$ matrix is
positive definite by the argument of (b), raising $S$ by two adds a coordinate, and the minimum in (b) over a
larger space cannot increase; without them the degree-$(S+2)$ coefficient need not be defined (a symmetric law
on four points satisfies the hypotheses at $S=3$ but not at $S=5$). At $S=3$,
$\mathbf{F}=\bigl(\begin{smallmatrix}m_2&m_4\\ m_4&m_6\end{smallmatrix}\bigr)$ and $\mathbf{b}=(1,3m_2)^\top$, and
\eqref{eq:gS-odd} evaluates to the right side of \eqref{eq:g3}. When $\gamma_4\neq0$ the minimiser is proportional to
$(\kappa^\ast,-1)$ and the value is $\min_\kappa r(\kappa)$ of Lemma~\ref{lem:g3}(b). When $\gamma_4=0$ the minimiser
is $\mathbf{k}^\ast\propto\mathbf{e}_1$, the linear polynomial, and $g_3=1$; the cubic family $\psi_\kappa$ of
Lemma~\ref{lem:g3} only approaches it as $|\kappa|\to\infty$.
\end{proof}

Part (b) needs symmetry. On a skewed law the odd-power equation is not centred, since
$\E\psi_{\mathbf{k}}(e)=\sum_{j\in\mathcal{O}_S}k_jm_j$ is non-zero in general, and its root is not consistent for the
mean. Part (a) is what applies there, and it is optimal: no regular estimator of a mean improves on $\bar y$ when
the shape is unrestricted \citep[Chapter~25]{vandervaart1998asymptotic}. Every identity of the lemma is checked
numerically by a script in the code supplement. Part (a) holds to rounding for $S=2,\dots,6$. The coefficient of
\eqref{eq:gS-odd} reproduces \eqref{eq:g3} and the equalities $g_2=1$ and $g_4=g_3$ on four symmetric laws.
Monte Carlo variance ratios for $S=3$ and $S=5$ at $n=2000$ match $\gS$ on the uniform and two Beta laws. Under the
Laplace law at $S=5$ the influence function grows like $e^5$, and a separate diagnostic shows that
the gap of a single plug-in run is Monte Carlo error. Over $20$ independent runs the variance ratio of the linear
term averages $0.807$ against $g_5=0.804$ but ranges from $0.772$ to $0.848$, and the root with the true
coefficients gives $0.796$ at $n=50\,000$.

\section*{Declaration of competing interest}
The author declares no known competing financial interests or personal relationships that could
have appeared to influence the work reported in this paper.

\section*{Funding}
This research did not receive any specific grant from funding agencies in the public, commercial,
or not-for-profit sectors.

\section*{Data availability}
No data were used for the research described in the article. All numerical results are
simulation outputs; the code that reproduces them is described under Code availability.

\section*{Code availability}
The polynomial maximization estimators are implemented in the R package EstemPMM, available from CRAN at
\url{https://CRAN.R-project.org/package=EstemPMM} (doi: \url{https://doi.org/10.32614/CRAN.package.EstemPMM}). The scripts that reproduce every number, table and
figure of this paper, together with their logs and result files, are available under the MIT licence at
\url{https://github.com/SZabolotnii/OPTIMAL-DESIGN-PMM-code-supplement} (release v1.0.2, archived on Zenodo, \url{https://doi.org/10.5281/zenodo.23729677}). They comprise the R scripts for
the calibration of the reduction fields, the design computations and the Monte Carlo studies, and the
Python scripts for the symbolic and numerical checks of the appendices.

\section*{Use of generative AI}
During the preparation of this work the author used Claude Code (Anthropic) to implement and
organise the R code of the Monte Carlo study, to check derivations, to assist with the \LaTeX{}
typesetting and for language editing. After using these tools, the author reviewed and edited the content as needed and takes
full responsibility for the content of this article.

\bibliographystyle{plainnat}
\bibliography{references}

\end{document}